\documentclass{article}

\usepackage{float}
\usepackage{lastpage}
\usepackage{titling}
\usepackage{authblk}
\usepackage{lipsum}

\usepackage[utf8]{inputenc}
\usepackage{amssymb}
\usepackage{amsmath}
\usepackage{graphicx}

\usepackage{helvet}

\usepackage{booktabs}
\usepackage{adjustbox}
\usepackage{threeparttable}
\usepackage{multirow}
\usepackage{adjustbox}

\usepackage[binary-units=true]{siunitx}
\DeclareSIUnit\pixel{pixel}
\DeclareSIUnit\fps{fps}

\newcommand*{\norm}[1]{\left\lVert#1\right\rVert}

\usepackage[super]{nth}

\usepackage{setspace}

\usepackage{hyperref}
\hypersetup{
	colorlinks=false,
	urlcolor=cyan,
}
\usepackage[capitalise,noabbrev]{cleveref}
\usepackage[numbers,sort&compress]{natbib}
\usepackage{titlesec}

\newcommand{\splitatcommas}[1]{%
	\begingroup
	\begingroup\lccode`~=`, \lowercase{\endgroup
		\edef~{\mathchar\the\mathcode`, \penalty0 \noexpand\hspace{0pt plus 1em}}%
	}\mathcode`,="8000 #1%
	\endgroup
}

\usepackage{appendix}
\crefname{appsec}{Appendix}{Appendices}
\Crefname{figure}{Fig.}{Figs.}

\title{Free-Breathing Liver Fat, $R_2^*$ and $B_0$ Field Mapping Using Multi-Echo Radial FLASH and Regularized Model-based Reconstruction}

\author[1,2]{Zhengguo Tan}
\author[3]{Christina Unterberg-Buchwald}
\author[1]{Moritz Blumenthal}
\author[1]{Nick Scholand}
\author[1]{Philip Schaten}
\author[1]{Christian Holme}
\author[1,2]{Xiaoqing Wang}
\author[4]{Dirk Raddatz}
\author[1,5]{Martin Uecker}

\affil[1]{Institute for Diagnostic and Interventional Radiology, University Medical Center G\"{o}ttingen, G\"{o}ttingen, Germany}
\affil[2]{Department of Artificial Intelligence in Biomedical Engineering, University of Erlangen–Nuremberg, Erlangen, Germany}
\affil[3]{Department of Cardiology and Pulmonology, University Medical Center G\"{o}ttingen, G\"{o}ttingen, Germany}
\affil[4]{Clinic of Gastroenterology, Gastrointestinal Oncology and Endocrinology, University Medical Center G\"{o}ttingen, G\"{o}ttingen, Germany}
\affil[5]{Institute of Biomedical Imaging, Graz University of Technology, Graz, Austria}

\begin{document}
\maketitle


\begin{abstract}
	\noindent
This work introduced a stack-of-radial multi-echo asymmetric-echo MRI sequence 
for free-breathing liver volumetric acquisition. 
Regularized model-based reconstruction was implemented 
in Berkeley Advanced Reconstruction Toolbox (BART) 
to jointly estimate all physical parameter maps 
(water, fat, $R_2^*$, and $B_0$ field inhomogeneity maps) 
and coil sensitivity maps from self-gated \textit{k}-space data. 
Specifically, locally low rank and temporal total variation regularization 
were employed directly on physical parameter maps. 
The proposed free-breathing radial technique 
was tested on a water/fat \& iron phantom, a young volunteer, 
and obesity/diabetes/hepatic steatosis patients. 
Quantitative fat fraction and $R_2^*$ accuracy were confirmed 
by comparing our technique with the reference breath-hold Cartesian scan.
The multi-echo radial sampling sequence achieves 
fast \textit{k}-space coverage and is robust to motion. 
Moreover, the proposed motion-resolved model-based reconstruction 
allows for free-breathing liver fat and $R_2^*$ quantification 
in multiple motion states. 
Overall, our proposed technique offers a convenient tool 
for non-invasive liver assessment with no breath holding requirement.
	
\end{abstract}

\section{Introduction}
\label{sec:introduction}

Quantitative parameter mapping of the liver is of interest 
in basic research and clinical practice. Specifically, quantitative 
proton density fat fraction (PDFF) and $R_2^*$ maps have been shown 
to be non-invasive imaging biomarkers for hepatic 
steatosis \cite{caussy_2018_fat,hu_2020_obesity} and 
iron overload \cite{wood_2011_iron,hernando_2014_iron}, respectively. 
Originating from the two-echo chemical-shift-encoded Dixon method 
\cite{dixon_1984_wf} for qualitative water/fat separation, 
quantitative assessment of liver fat and iron decomposition firstly requires 
multi-echo chemical-shift encoding (e.g.~low flip angle multi-gradient-echo acquisition). 
Conventional acquisition methods require subjects to hold their breath.
Therefore, incomplete breath hold or incompliant patients can induce 
pronounced image artifacts, 
thereby hampering the quantification of liver fat and $R_2^*$.

To address the respiratory motion problem, 
several recent works proposed free-breathing liver fat and $R_2^*$ mapping. 
Armstrong et al.~\cite{armstrong_2018_fat} and 
Zhong et al.~\cite{zhong_2020_r2sclinic,zhong_2020_resr2s} 
proposed to use stack-of-radial multi-echo sampling with bipolar gradients. 
Sampled echoes were binned into four respiratory phases 
and reconstructed via non-uniform FFT (NUFFT) \cite{fessler_2003_nufft}. 
Liver fat and $R_2^*$ maps are quantified 
via physics modeling \cite{yu_2007_t2sideal,yu_2008_mft2sideal,chebrolu_2010_indiwf} 
and image-space fitting \cite{reeder_2005_ideal,hernando_2010_gc,zhong_2014_wfadafit}. 
Schneider et al.~\cite{schneider_2020_mobawfr2s} 
employed model-based reconstruction 
\cite{block_2009_mobat2,fessler_2010_moba,doneva_2010_mobawf} 
to jointly reconstruct water, fat, and $R_2^*$ maps directly from $k$-space data 
with spatial and temporal total variation (TV) regularization, 
whereas the $B_0$ field inhomogeneity map was pre-calibrated \cite{hernando_2010_gc} 
and was then kept fixed during iterative reconstruction. 
Wang et al.~\cite{wang_2022_mt-me} implemented 
inversion recovery (IR) magnetization preparation before multi-echo readouts 
and applied the multi-tasking reconstruction technique \cite{christodoulou_2018_mt} 
to reconstruct respiratory motion resolved multi-echo and IR images. 
Subsequently, these images were used for model fitting \cite{hernando_2010_gc} 
to obtain quantitative fat, $R_2^*$, and water-specific $T_1$ maps. 
Starekova et al.~\cite{starekova_2022_fb-fat} proposed a 2D acquisition scheme 
with a non-local means \cite{buades_2005_nlm} motion-corrected averaging technique. 

Among these free-breathing techniques, only the work from Schneider et al. employed 
model-based reconstruction. However, this work calibrated sensitivities
and field map based on an a fast initial reconstruction with lower quality.
This is suboptimal, as these maps are then either affected by inconsistencies due 
to lower temporal resolution when averaging data from multiple motion states
or affected by undersampling artefacts.
Joint estimation of the field map can reduce errors. It was proposed 
before using a smoothness prior on the $B_0$ field map via first- or 
second-order finite-difference regularization 
\cite{sutton_2004_dynamicfield,olafsson_2008_joint,funai_2008_secondorder},
Integration of the field map into a motion-resolved reconstruction for
free-breathing liver imaging poses substantial additional challenges,
as the $B_0$ field map appears as a non-linear term in the water/fat separation
and may change rapidely at water-tissue interfaces.

The contribution of this work is two-fold:
First, to achieve more efficient and faster acquisition, 
our work combined multi-echo asymmetric-echo radial fast low-angle shot (FLASH) 
with stack-of-stars volumetric acquisition \cite{block_2014_rad}
for free-breathing acquisition. 
Second, we implemented a regularized model-based reconstruction method 
for joint estimation of all quantitative parameter maps as well as
$B_0$ map, and coil sensitivity maps directly from $k$-space data,
thus extending the approach used by Schneider et al.
We solved the nonlinear inverse problem via 
iteratively regularized Gauss-Newton method (IRGNM) \cite{uecker_2008_nlinv} 
with the alternating direction method of multipliers (ADMM), 
allowing for generalized regularization terms \cite{boyd_2010_admm}. 
We validated the proposed acquisition and reconstruction methods by 
comparing with the reference breath-hold Cartesian scan \cite{zhong_2014_wfadafit} 
in volunteers and patients diagnosed with obesity, diabetes, or non-alcoholic fatty liver disease (NAFLD).

\section{Theory}

\subsection{Multi-Echo Radial Sampling}

\begin{figure}
	\centering
	\includegraphics[width=\columnwidth]{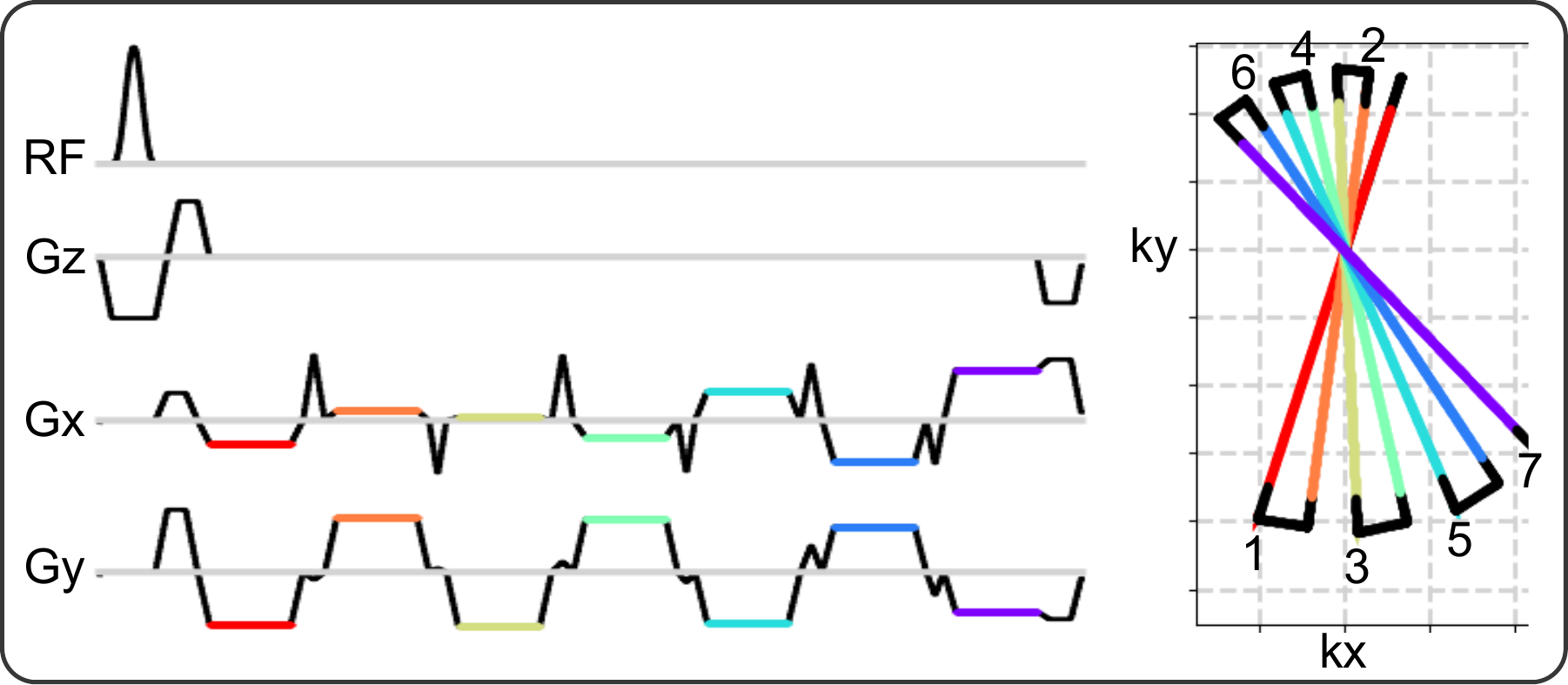}
	\caption{(Left) One representative repetition time (TR) block of 
		the proposed multi-echo asymmetric-echo radial sequence. 
		(Right) The corresponding \textit{k}-space trajectory. 
		The echoes are color coded, indicating the period when ADC is switched on, 
		while the dark solid lines indicate either the ramp or the blip gradients.}
	\label{FIG:SEQ}
\end{figure}

Data acquisition is based on the multi-echo radial sampling sequence 
\cite{tan_2019_mobawf}. 
As depicted in \cref{FIG:SEQ}, seven gradient echoes per RF excitation 
were acquired. 
All radial spokes within one frame were uniformly distributed in \textit{k}-space 
and the angle increment between frames was the small Golden angle 
\cite{winkelmann_2007_golden} ($\approx 68.75^o$).
The angles of the radial spokes within one frame are 
$\theta_{l,m} = 2\pi \cdot [ (l-1) \cdot N_\text{echo} + m - 1 ] / (N_\text{echo} \cdot N_\text{shot})$ 
with $l$ and $m$ being the excitation and echo index, respectively. 
To shorten TE, TR as well as scan time, 
asymmetric echoes (i.e.~partial Fourier readouts) \cite{untenberger_2016_asym} 
were employed in combination with multi-echo sampling.

\subsection{Nonlinear Signal Model}

Parallel MRI \cite{roemer_1990_pi,pruessmann_1999_sense,griswold_2002_grappa} 
simultaneously receives signals from multiple receiver coils, 
and is extendable to include multiple echoes when using long echo-train MRI sequences,
\begin{equation} \label{EQU:y_pimeco}
	y_{j,m}(t) = \int \text{d} \vec{r} \, e^{-i 2\pi \vec{k} (t) \cdot \vec{r} } c_j(\vec{r}) \rho_m(\vec{r}) \;,
\end{equation}
with $c_j$ and $\rho_m$ being the $j$th coil sensitivity map and the $m$th echo image, respectively. 
$y_{j,m}(t)$ is the acquired multi-coil multi-echo \textit{k}-space data.
In the case of gradient echoes, $\rho_m$ is governed by
\begin{equation} \label{EQU:meco_mspec}
	\rho_m = \bigg( \sum_i I_i \cdot e^{i2\pi f_i \text{TE}_m} \cdot e^{-{R_2^*}_i \text{TE}_m} \bigg) \cdot e^{i2\pi f_{B_0} \text{TE}_m} \; ,
\end{equation}
where the first term sums up signals from all chemical species (indexed by $i$), 
characterized by their corresponding proton density ($I_i$), 
resonance frequency ($f_i$) and relaxation rate (${R_2^*}_i$). 
Here, the dependency on the spatial coordinates $\vec{r}$ is suppressed for simplicity.
In addition, the echo signal is modulated by the $B_0$ field inhomogeneity. 
$\text{TE}_m$ denotes the $m$th echo time.

This generalized multi-species signal can be simplified to only two compartments \cite{yu_2007_t2sideal,yu_2008_mft2sideal,chebrolu_2010_indiwf}, 
i.e.~water (W) and fat (F), 
\begin{equation} \label{EQU:meco_wfr2s}
	\rho_m = \bigg( \text{W} + \text{F} \cdot z_m \bigg) \cdot e^{- R_2^* \text{TE}_m} \cdot e^{i2\pi f_{B_0} \text{TE}_m} \; .
\end{equation}
The chemical-shift phase modulation from fat is denoted as 
$z_m$ with the 6-peak fat spectrum \cite{yu_2008_mft2sideal}, 
while all fat peaks are assumed to have an equal $R_2^*$ \cite{reeder_2012_perfr2s}.
$\text{W}$ and $\text{F}$ are complex-valued, 
while $R_2^*$ and $f_{B_0}$ are real. 

Given the above MR signal model, the nonlinear forward model in multi-coil multi-echo acquisition 
can be written in the operator form 
\begin{equation} \label{EQU:op_fwd}
	y_{j,m} = F_{j,m} (x) := P_m \mathcal{F} M \mathcal{S} \mathcal{B} \; ,
\end{equation}
with $x = (\text{W}, \text{F}, R_2^*, f_{B_0}, \cdots, c_j, \cdots)^T$. 
$F_{j,m} (x)$ denotes the forward operator.
$j$ is the the coil index ($j \in [1,N_\text{coil}]$), 
and $m$ the echo index ($m \in [1, N_\text{echo}]$). 
The nonlinear operator ($\mathcal{B}$) calculates echo images 
according to the parameter maps in $x$ and 
the corresponding signal model as given in \cref{EQU:meco_wfr2s}. 
Every echo image is then pointwisely multiplied by every coil sensitivity map in $x$, 
as denoted by the operator $\mathcal{S}$. 
Afterward, all multi-echo coil images are masked to be restricted to a given
field of view (FOV) ($M$), Fourier-transformed ($\mathcal{F}$), and sampled ($P$) at each echo.
In addition, because radial sampling acquires \textit{k}-space data within a circular region, 
the \textit{k}-space filter \cite{pruessmann_2001_gsense}
is applied to the sampling pattern to suppress potential checker-board artifacts.

\subsection{Model-based Nonlinear Inverse Reconstruction}

The joint estimation of the unknown $x$ is a nonlinear inverse problem,
\begin{equation} \label{EQU:obj_func}
	\begin{aligned}
		\text{minimize}~& \sum_t \sum_m \sum_j \norm{y_{j,m,t} - F_{j,m,t}(x)}_2^2 \\
		& + \lambda_1 \norm{\mathcal{L} (E_1 x)}_1 + \lambda_2 \norm{\text{TV}_t (E_2 x)}_1 \\
		& + \lambda_3 [ \norm{x}_2^2 + \norm{T_{f_{B_0}} f_{B_0}}_2^2 + \norm{T_C C}_2^2 ]\\
		\text{subject to}~& {R_2^*} \geq 0
	\end{aligned}
\end{equation}
The unknown $x$ in this problem contains respiratory-motion-resolved 
model parameter maps ($\text{W}$,  $\text{F}$, $R_2^*$, and $f_{B_0}$) 
as well as coil sensitivity maps, and thus has the shape of 
$[N, N, 4 + N_\text{coil}, N_\text{bin}]$. 
Here, $[N, N]$ corresponds to the image matrix size, 
$4$ is the total number of model parameters, 
and $N_\text{bin}$ is the number of respiratory bins.

The locally low rank (LLR) regularization $\mathcal{L}$ with the regularization strength $\lambda_1$
was applied onto the parameters $\mathrm{W}$, $\mathrm{F}$ and $R_2^*$. 
These parameter maps were extracted from the complex $x$ matrix using the extraction operator $E_1$. 
The block size of the LLR constraint was $16 \times 16$ 
with random shifts of the blocks among iterations.

Temporal TV regularization \cite{feng_2014_grasp} 
with the regularization strength $\lambda_2$ was applied onto 
all model parameter maps ($\mathrm{W}$, $\mathrm{F}$, $R_2^*$, and $f_{B_0}$), 
which can be extracted via the operator $E_2$.

In addition, $\ell2$ regularization with the regularization strength $\lambda_3$ 
was applied to all unknowns. 
To enforce spatial smoothness of the $B_0$ and coil sensitivity maps, 
the Sobolev-norm weighting \cite{uecker_2008_nlinv} 
was utilized to penalize high spatial frequencies, 
\begin{equation}
	T = \mathcal{F}^{-1} \bigg(1 + w \cdot \norm{\vec{k}} \bigg)^{-h} \mathcal{F}
\end{equation}
$h$ is set as $16$ for both maps, 
while $w_{f_{B_0}} = 22$ and $w_{C} = 220$ for the $f_{B_0}$ and coil sensitivity maps, 
respectively. $\vec{k}$ is a 2D Cartesian grid ranging from $-0.5$ to $0.5$ 
\cite{tan_2019_mobawf}.

The objective functional in \cref{EQU:obj_func} was solved by IRGNM with ADMM. 
For details about this algorithm please refer to 
\cref{SEC:APPENDIXA,SEC:APPENDIXB}. 
Our implementation enables flexible selections of regularization terms. 

We compared our proposed regularization strategy against only $\ell^2$ regularization 
(i.e.~removal of the $\ell^1$ terms in \cref{EQU:obj_func}) 
as well as the temporal TV regularization method 
without $B_0$ update (similar to the work from Schneider et al.~\cite{schneider_2020_mobawfr2s}). 
Keeping $B_0$ constant during model-based reconstruction can be realized via 
setting the forward model's adjoint of derivative term with respect to $B_0$ as 0 
(Previously, this has been applied to nonlinear inversion as parallel imaging to 
keep coil sensitivity maps constant during iterative reconstruction \cite{uecker_2008_nlinv}).

\section{Methods}

\subsection{Phantom Experiments}

\begin{figure}
	\centering
	\includegraphics[width=\columnwidth]{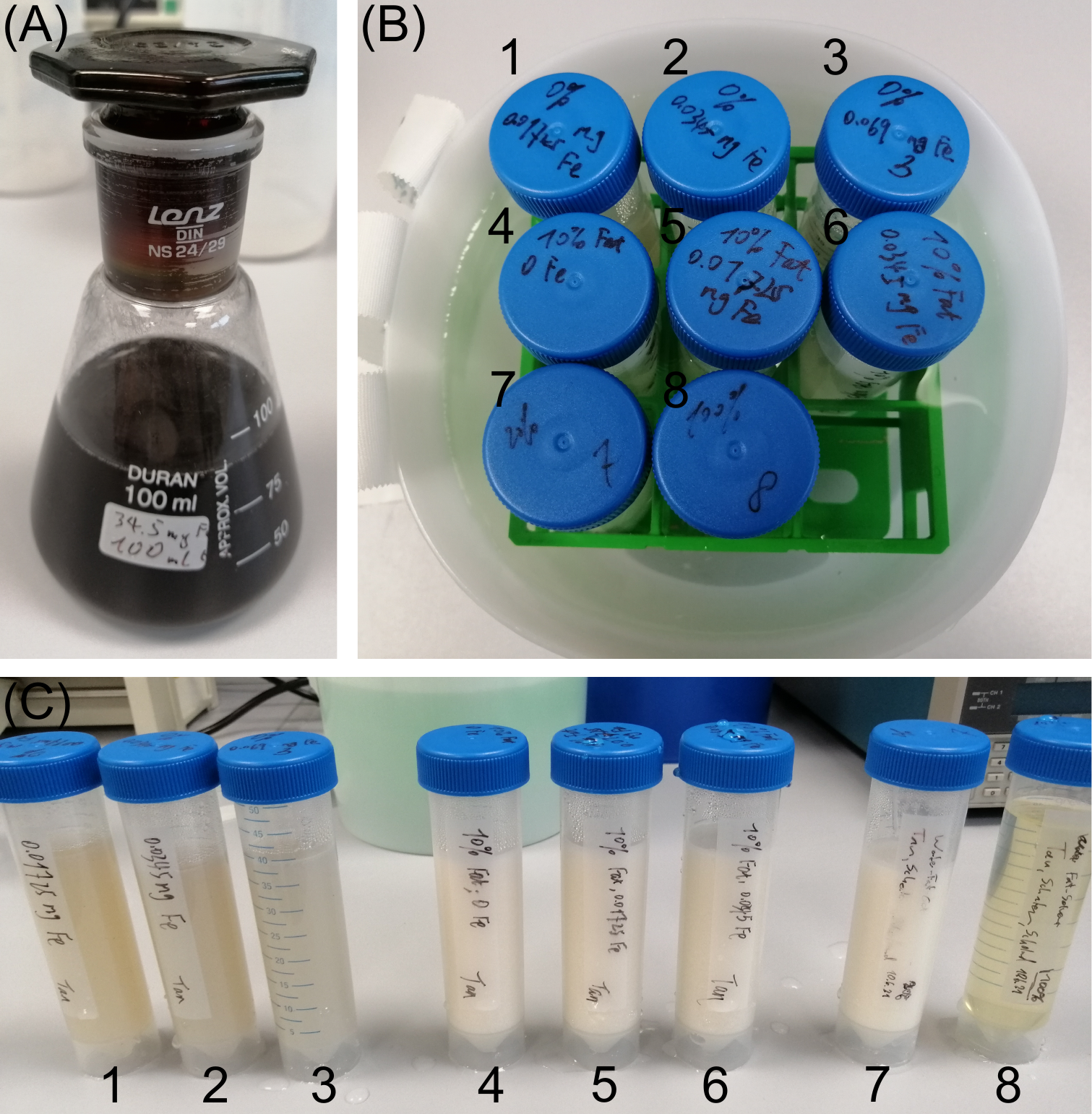}
	\caption{Photos of the constructed water/fat \& iron phantom. 
		(A) 34.5~mg iron nano particle diluted in 100~mL distilled water.
		(B) Phantom layout with eight tubes as listed in (C).}
	\label{FIG:PHA_WFIRON}
\end{figure}

\begin{table*}[t]
	\caption{Measured volume of iron, water and peanut oil solutions for the eight tubes}
	\setlength{\tabcolsep}{3pt}
	\begin{tabular}{m{0.1\textwidth} m{0.1\textwidth} m{0.1\textwidth} m{0.1\textwidth} m{0.1\textwidth} m{0.1\textwidth} m{0.1\textwidth} m{0.1\textwidth} m{0.1\textwidth}}
		\toprule
		Solution (\si{\milli\liter}) & Tube 1 & Tube 2 & Tube 3 & Tube 4 & Tube 5 & Tube 6 & Tube 7 & Tube 8 \\
		\hline
		Iron  &  0.5 &  1.0 &  2.0 &  0 &  0.5 &  1.0 &  0 &  0 \\
		Water & 50   & 50   & 50   & 45 & 45   & 45   & 40 &  0 \\
		Fat   &  0   &  0   &  0   &  5 &  5   &  5   & 10 & 50 \\
		\bottomrule
	\end{tabular}
	\label{TAB:PHA}
\end{table*}

A water/fat \& iron phantom \cite{hines_2009_wfiron,bush_2018_fat} was constructed. 
Details of this phantom are provided in \cref{FIG:PHA_WFIRON} and \cref{TAB:PHA}. 
Due to water lost as steam in the preparation of the water and fat emulsion, 
the actual fat fraction was higher than prescribed in \cref{TAB:PHA}. 
Therefore, the actual fat fraction values of every tube were validated using 
the standard Siemens MR spectroscopy (MRS) protocol.
MRI experiments were conducted with an 18-channel body matrix coil 
together with a spine coil on 
\SI{3}{\tesla} (Skyra, Siemens Healthineers, Erlangen, Germany). 
The proposed multi-echo radial sampling sequence was used 
with the following parameters:~FOV \SI{200}{\milli\metre}, 
base resolution 160, spatial resolution 
\SI{1.25}{\milli\meter}~$\times$~\SI{1.25}{\milli\meter}~$\times$~\SI{3.5}{\milli\meter},
bandwidth \SI{1040}{\hertz/\pixel}, flip angle \SI{4}{\degree}, 
asymmetric echo \cite{untenberger_2016_asym} of \SI{30}{\percent}, 
TE $\splitatcommas{1.07, 2.35, 3.23, 4.51, 5.39, 6.67, 7.55}$~\si{\ms} 
and TR $\SI{8.81}{\ms}$.

\subsection{In Vivo Experiments}

For in vivo scans, the same coils and MRI system were used. 
Two radial protocols were implemented, one with a smaller field of view (FOV) 
\SI{320}{\milli\meter} for subjects with relatively small body size (Volunteer \#1), 
and another with a larger FOV \SI{410}{\milli\meter} 
which fits well for obese patients (Patient \#1 to \#9). 
Breath-hold multi-echo Cartesian Dixon MRI was used as the reference. 
Detailed parameters are provided in \cref{TAB:Protocol}. 
Although the Cartesian scan requires one single breath hold 
and is fast, it captures only one motion state. 
In contrast, the proposed radial scan allows for free breathing, 
but requires longer scan time (2:47~\si{\minute}). 
We therefore, also investigated the effectiveness of 
the proposed motion-resolved model-based reconstruction 
on retrospectively undersampled (shortend)
radial data corresponding to a scan time of 1:24~\si{\minute}.

\begin{table}
	\begin{threeparttable}
		\caption{Imaging parameters for (left) the breath-hold Cartesian reference 
			and (right) the proposed free-breathing multi-echo radial acquisition 
			with two different FOV}
		\label{TAB:Protocol}
		\setlength{\tabcolsep}{3pt}
		\begin{tabular}{m{0.36\columnwidth} m{0.18\columnwidth} m{0.18\columnwidth} m{0.18\columnwidth}}
			\toprule
			& Cartesian & Radial & Radial \\
			\hline
			Flip angle ($^o$) & $4$ & $4$ & $4$ \\
			Bandwidth (Hz/pixel) & 1080 & 1090 & 1090 \\
			Asymmetric echo & None & \SI{30}{\percent} & \SI{30}{\percent} \\
			Number of echoes & 6\tnote{a} & 7\tnote{b} & 7\tnote{c} \\
			Repetition Time (\si{\milli\second}) & $9$ & $8.35$ & $8.39$ \\
			FOV (mm~$\times$~mm) & $410 \times 348$ & $320 \times 320$ & $410 \times 410$ \\
			Pixel size (mm~$\times$~mm) & $2.56 \times 2.56$ & $1.6 \times 1.6$ & $1.6 \times 1.6$ \\
			Base resolution & $160$ & $200$ & $256$ \\
			Slice thickness (\si{\milli\meter}) & $3.5$ & $3.5$ & $3.5$ \\
			Slice resolution (\si{\percent}) & $50$ & $100$ & $100$ \\
			Slice oversampling (\si{\percent}) & 25 & 25 & 25 \\
			Number of slices & 64 & 48 & 48 \\
			Scan time (min:sec) & 0:20 & 2:45 & 2:47 \\
			\bottomrule
		\end{tabular}
		\begin{tablenotes}
			\item[a] Echo times are $1.06, 2.46, 3.69, 4.92, 6.15, 7.38$~ms;
			\item[b] Echo times are $0.98, 2.20, 3.04, 4.26, 5.10, 6.32, 7.16$~ms;
			\item[c] Echo times are $0.98, 2.21, 3.05, 4.28, 5.12, 6.35, 7.19$~ms.
		\end{tablenotes}
	\end{threeparttable}
\end{table}

One volunteer and twenty obesity/diabetes/NAFLD patients were scanned
with both the multi-echo radial sampling sequence and the Cartesian
DIXON sequence. The data for five patients were excluded because the
automatic calibration of the resonance frequency used was the fat
instead of the water peak, which then causes the reconstruction to fail.
While this can be corrected in post-processing, this is not yet
implemented.
All subjects gave written informed consent before MRI 
in compliance with the regulations established by the local ethics committee. 
A summary of all subjects is given in \cref{TAB:Subject}, 
including the age, gender, height, weight, and the body mass index (BMI).

\begin{table}
		\caption{Summary of Subjects Participated in the Study}
		\label{TAB:Subject}
		\setlength{\tabcolsep}{3pt}
		\begin{tabular}{m{0.17\columnwidth} m{0.06\columnwidth} m{0.08\columnwidth} m{0.17\columnwidth} m{0.17\columnwidth} m{0.20\columnwidth}}
			\toprule
			& Age & Gender & Height (\si{\centi\meter}) & Weight (\si{\kilogram}) & BMI (\si{\kilogram/\square\meter}) \\
			\hline
			Volunteer \#1 & 27 & F & 155 &  60 & 25.0 \\
			Patient \#1   & 55 & M & 184 & 105 & 31.0 \\ 
			Patient \#2   & 60 & F & 172 & 105 & 35.5 \\ 
			Patient \#3   & 64 & F & 164 &  80 & 29.7 \\ 
			Patient \#4   & 60 & F & 153 & 112 & 47.8 \\ 
			Patient \#5   & 44 & F & 165 &  94 & 34.5 \\ 
			Patient \#6   & 43 & F & 165 & 165 & 60.6 \\ 
			Patient \#7   & 65 & M & 176 &  98 & 31.6 \\ 
			Patient \#8   & 55 & F & 156 &  88 & 36.2 \\ 
			Patient \#9   & 54 & F & 160 &  68 & 26.6 \\ 
			
			Patient \#10  & 62 & M & 202 & 100 & 24.5 \\ 
			Patient \#11  & 67 & M & 176 &  99 & 32.0 \\ 
			Patient \#12  & 26 & F & 178 & 180 & 56.8 \\ 
			Patient \#13  & 69 & M & 191 & 134 & 36.7 \\ 
			Patient \#14  & 61 & M & 173 & 114 & 38.1 \\ 
			Patient \#15  & 51 & M & 169 & 101 & 35.4 \\ 
			\bottomrule
		\end{tabular}
\end{table}

\subsection{Model-based Reconstruction}

In the spirit of reproducible and open research, 
the proposed regularized model-based reconstruction is made
publicly available in BART \cite{uecker_2015_bart}. 
Scripts to produce the experiments are available at \url{https://github.com/mrirecon/multi-echo-liver} upon publication. 
A brief description of the reconstruction procedure is given here.

\subsubsection*{Pre-Processing}

The acquired multi-coil multi-echo data was compressed to ten virtual coils 
via principal component analysis \cite{huang_2008_scc}. 
The multi-echo sampling trajectory 
was corrected for gradient delays using the
radial spoke intersections for gradient delay estimation (RING) method 
\cite{rosenzweig_2019_ring}. 
RING determines the intersection points of different radial spokes. 
Afterward, RING estimates the gradient delay coefficients 
given the position of the intersection points. 
RING was applied to every echo to estimate 
its corresponding gradient delay coefficients, 
which were then used to correct the trajectories.

\subsubsection*{Binning}

In this work, the singular spectrum analysis (SSA-FARY) technique 
\cite{rosenzweig_2020_ssa} 
was adapted for self-gating of respiratory motions. 
SSA-FARY combines the ideas of time-delayed embedding and principal component analysis, 
and allows for robust extraction of oscillatory motion signals 
such as respiration from an auto-calibration (AC) region. 
In particular, for a periodic signal contained in the AC region, 
SSA-FARY yields a quadrature pair, which can be used for self-gating. 
Such a quadrature pair can be thought of as a generalized sine-cosine-pair 
describing the respective periodic motion and can be represented by a phase-portrait, 
where an angle of $0^o$ states the beginning and $360^o$ states the end of a motion cycle. 

In this work, we chose to split the respiratory motion into $N = 4$ bins, 
hence, each bin represents a circular sector with a central angle of $360^o / N$. 
Then, the respiratory motion state at a given time is determined by 
the respective angle defined by the quadrature pair.
As one echo train was relatively short, 
only the first echo was extracted for SSA-FARY.

Note that although more respiratory bins can improve the temporal resolution, 
it leads to fewer radial spokes per bin. Such a trade-off can be alleviated 
via the temporal TV regularization. However, we observed that increasing bins 
did not significantly affect the reconstructed image quality (results not shown). 
Since the reference Cartesian scan was performed during end expiration, 
we used the end-expiration bin of the radial scan for comparison.

\subsubsection*{Initialization}

Off-resonance phase modulation causes phase wrapping along the echoes, 
especially in the case of large $B_0$ field inhomogeneity. 
Consequently, multiple local minima could occur for the field map $f_{B_0}$. 
To prevent this, the $f_{B_0}$ map was initialized 
from a model-based three-point water/fat separation \cite{tan_2019_mobawf}. 
$\mathrm{W}$ and $\mathrm{F}$ were initialized as \num{0.1}. 
${R_2^*}$ and coil sensitivity maps were initialized as \num{0}.

\subsubsection*{Iterative Reconstruction}

For the model-based reconstruction, the regularization strength in \cref{EQU:obj_func} 
(e.g.~$\lambda_1$, $\lambda_2$, and $\lambda_3$) 
is reduced along Newton steps, e.g.~$\lambda_i^{(n)} = \lambda_i / D^{n-1}$ for $\lambda_1$, 
with $n$ being the $n$th Newton iteration 
and the reduction factor $D > 1$.
$D = 3$ was used in this work. 
For ADMM, the maximum number of iterations in each Newton step 
were given as:~$n_\text{maxiter} = \min (M, 10 \times 2^{- \ln \lambda^{(n)}})$, 
where $M$ was set as \num{200}. 
Consequently, the maximal iterations gradually increase along the Newton steps.

In this work, eight Newton steps were employed, 
and the ADMM penalty parameter $\rho$ was set as $0.001$. 
For the regularization terms in \cref{EQU:obj_func}, we set 
$\lambda_1 = 0.0003$, $\lambda_2 = 0.1$, and $\lambda_3 = 1$. 
The scaling of the F, $R_2^*$, and $B_0$ maps was set as 
$1.6$, $0.001$, and $1$, respectively.
All reconstructions were performed on a Tesla V100-SXM2 32~GB GPU 
(NVIDIA, Santa Clara, CA, USA).

\subsubsection*{Post-Processing}

With the reconstructed water and fat images, fat fraction (FF) maps can be computed 
taking into account the magnitude discrimination \cite{liu_2007_ff}, 
i.e.~$\text{FF} = 1 - |\text{W}| / (|\text{W}| + |\text{F}|)$ 
for pixels in which water dominates.

\section{Results}

\begin{figure}
	\centering
	\includegraphics[width=\columnwidth]{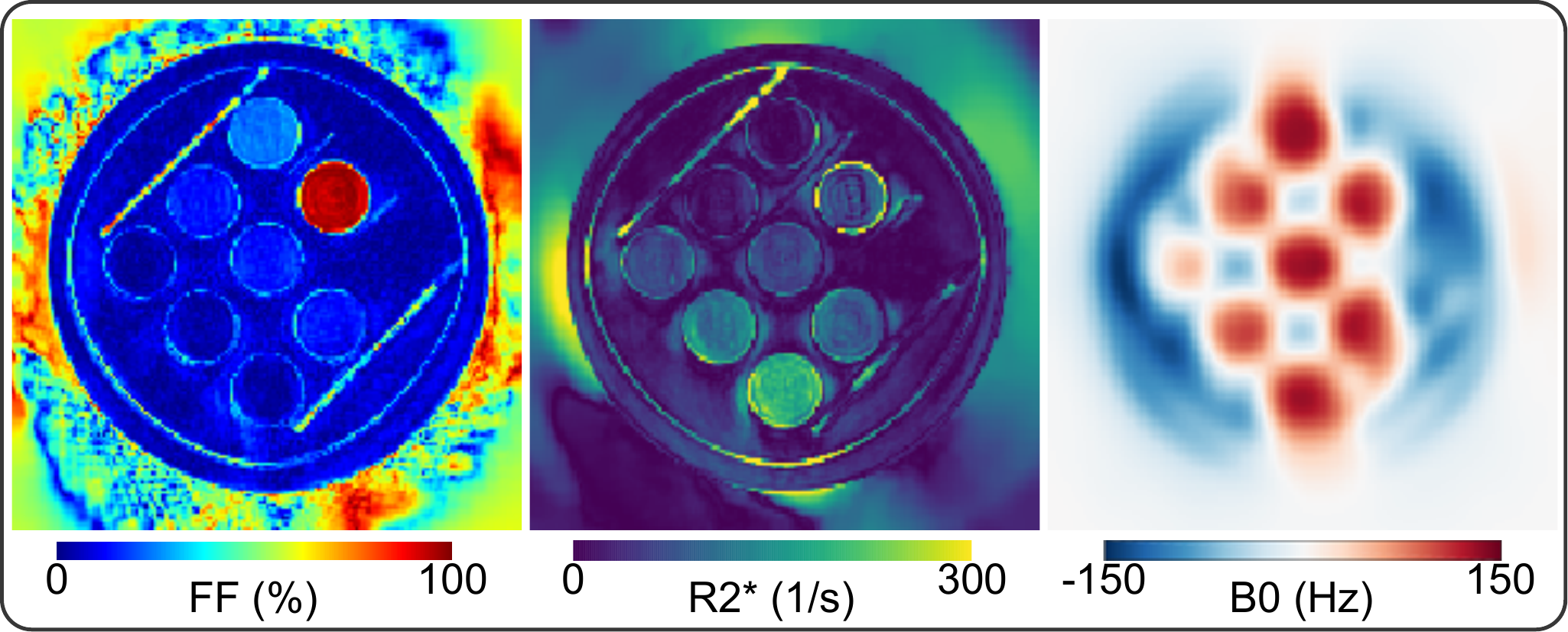}
	\caption{Multi-echo radial FLASH acquisition and 
		model-based reconstruction results 
		of the water/fat \& iron phantom built in-house. 
		Displayed images are FF, $R_2^*$, and $B_0$ maps, respectively.}
	\label{FIG:PHA}
\end{figure}

\begin{figure}
	\centering
	\includegraphics[width=\columnwidth]{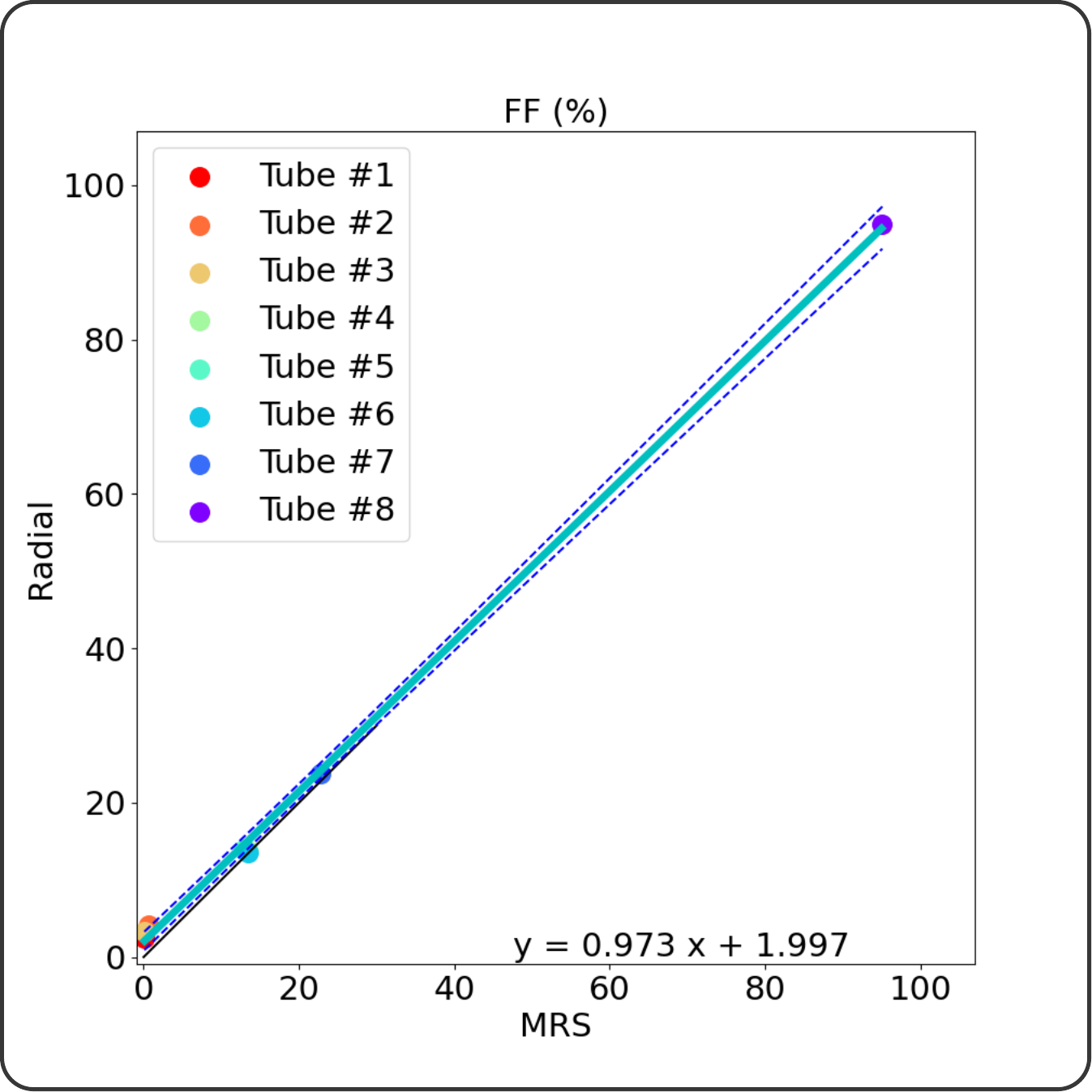}
	\caption{Quantitative analysis of FF values 
		for the eights tubes via linear regression 
		between the standard MR Spectroscopy (MRS) 
		and the proposed multi-echo radial acquisition 
		with model-based reconstruction.}
	\label{FIG:PHA_LINFIT}
\end{figure}

With the proposed multi-echo radial FLASH acquisition and 
model-based reconstruction, \cref{FIG:PHA} shows the results 
of the in-house built water/fat \& iron phantom. 
The FF values of every tube are validated against 
the MRS measurements. 
The FF values of all tubes match well between the two measurements, 
as shown in \cref{FIG:PHA_LINFIT}.

\begin{figure*}
	\centering
	\includegraphics[width=\textwidth]{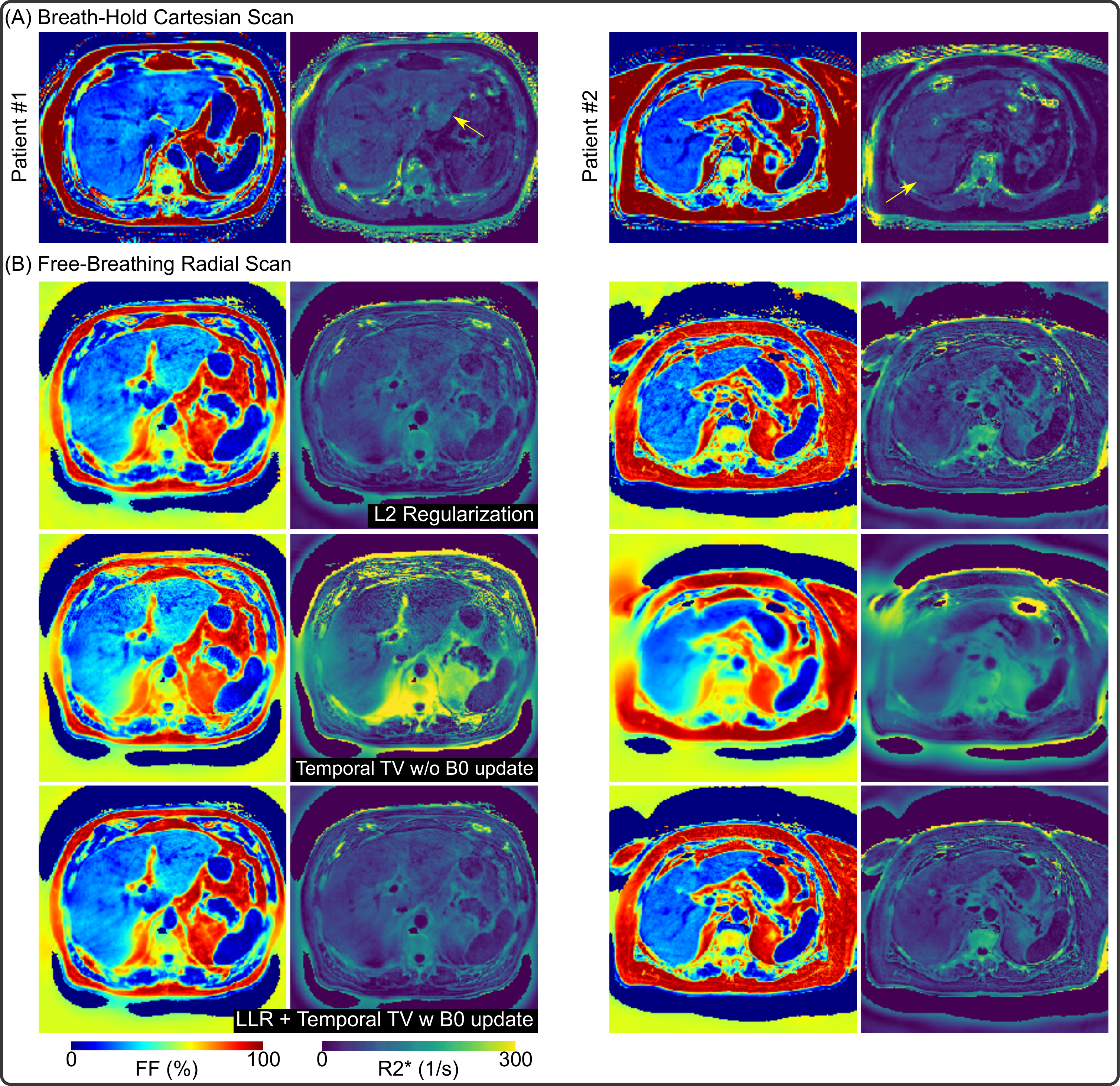}
	\caption{Comparison of (A) the reference breath-hold Cartesian scan and 
		(B) the proposed free-breathing radial scan on Patients \#1 and \#2. 
		Furthermore, for the radial data in (B), 
		we compared three different regularizations: 
		L2 regularization, 
		temporal TV regularization without $B_0$ update, 
		and the proposed spatial LLR and temporal TV regularization with $B_0$ update.}
	\label{FIG:REGU}
\end{figure*}

\begin{figure*}
	\centering
	\includegraphics[width=\textwidth]{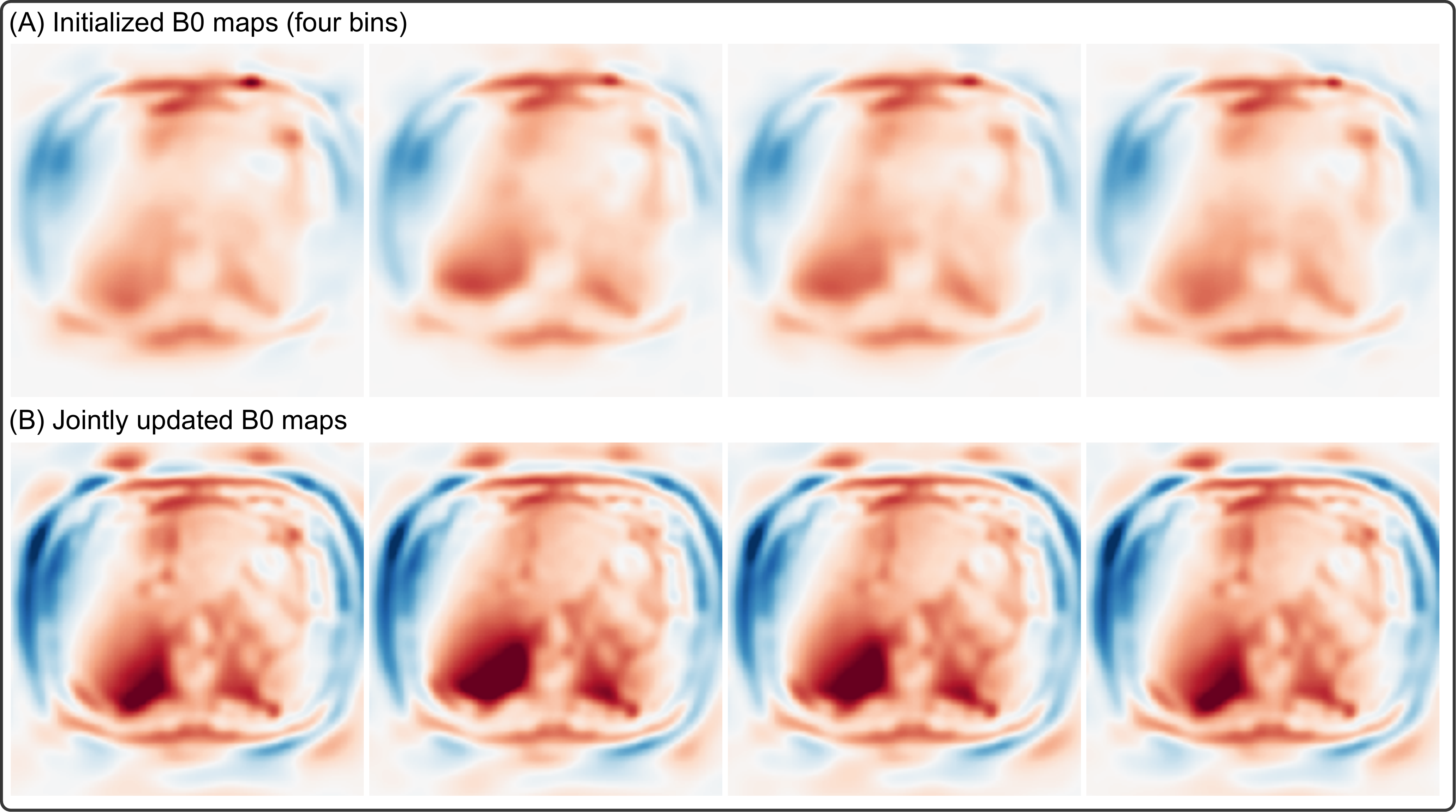}
	\caption{(A) Initial $B_0$ maps of four respiratory bins 
		for the model-based reconstruction without $B_0$ update in \cref{FIG:REGU}.
		(B) Jointly updated $B_0$ maps of four respiratory bins. 
		Joint update using seven echoes helps to recover more details in 
		the $B_0$ maps.}
	\label{FIG:B0}
\end{figure*}

\Cref{FIG:REGU} shows the reconstructed FF and $R_2^*$ maps of Patients \#1 and \#2
from the reference breath-hold Cartesian and the proposed free-breathing radial scan, 
respectively. 
The $R_2^*$ maps from Cartesian scans suffer from noise in the middle region 
(likely due to undersampling in the phase-encoding direction). 
Moreover, as pointed out by the yellow arrows, 
incomplete breath hold results in artifacts in the $R_2^*$ maps.
For model-based reconstruction of the parameter
maps ($\mathrm{W}$, $\mathrm{F}$, $R_2^*$ and $f_{B_0}$) 
from the radial scan, three different versions were compared: 
$\ell^2$, temporal TV without $B_0$ update 
(as employed by Schneider et al.~\cite{schneider_2020_mobawfr2s}), 
LLR plus temporal TV with $B_0$ update. 
The results show that residual streaking artifacts can be suppressed 
via the use of spatial LLR and temporal TV regularization.
An update of the $B_0$ map during the model-based iterative reconstruction
improves the reconstructed FF and $R_2^*$ maps. The current initialization
strategy used 
only the first three echoes with limited coverage of $k$-space, 
which supplies only a rough estimate of the $B_0$ map.
Performing $B_0$ updates can help to recover a more accurate $B_0$ map
by using information from all seven echoes as shown in \cref{FIG:B0}.
All subsequent figures use 
the same color maps for FF and $R_2^*$ as in \cref{FIG:REGU}.

\begin{figure*}
	\centering 
	\includegraphics[width=\textwidth]{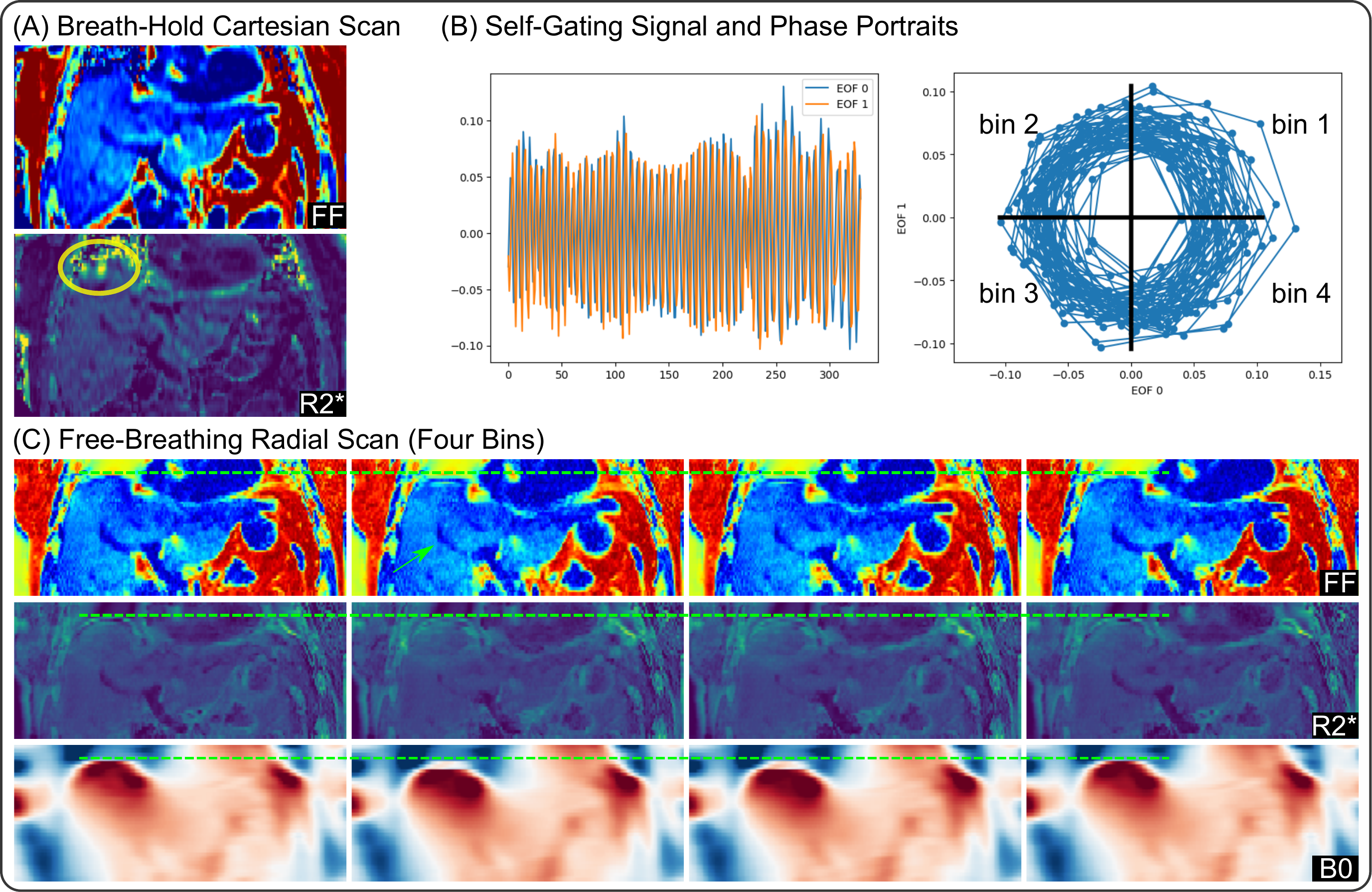}
	\caption{(A) One coronal-view FF and $R_2^*$ map 
		from breath-hold Cartesian scan of Patient \#1. 
		(B) Plots of self-gating signal and phase portraits. 
		(C) Four coronal-view FF and $R_2^*$ maps from 
		free-breathing radial scan with self-gated motion-resolved 
		model-based reconstruction. 
		The $R_2^*$ map in (A) suffers from 
		hyper intensities in the liver dome region (yellow ellipse), 
		whereas the proposed radial scan shows consistent $R_2^*$ values.}
	\label{FIG:SSA}
\end{figure*}

Further, to demonstrate the effectiveness of using SSA-FARY for 
the motion resolved free-breathing radial scan, 
\cref{FIG:SSA} compares the reference breath-hold Cartesian scan 
and the proposed free-breathing radial scan 
for Patient \#1 in the reformatted coronal view. 
While the liver dome region shows hyper $R_2^*$ intensity from the Cartesian scan, 
consistent $R_2^*$ values are visible 
from the proposed radial scan and model-based reconstruction. 
Such hyper intensity may be caused by the fast $B_0$ field change 
in the liver dome region, which is close to the lung with much lower MR signal. 
On the other hand, respiratory motion is well separated across the four frames. 
As shown in the self-gating AC signal plot in \cref{FIG:SSA} (B), 
A total of 330 radial spokes were acquired during the free-breathing scan. 
When plotted into the phase portrait, 
the pair of empirical orthogonal functions (EOF) can be equally divided into 
four angular blocks, representing four distinct respiratory motion states.

\begin{figure*}
	\centering
	\includegraphics[width=0.75\textwidth]{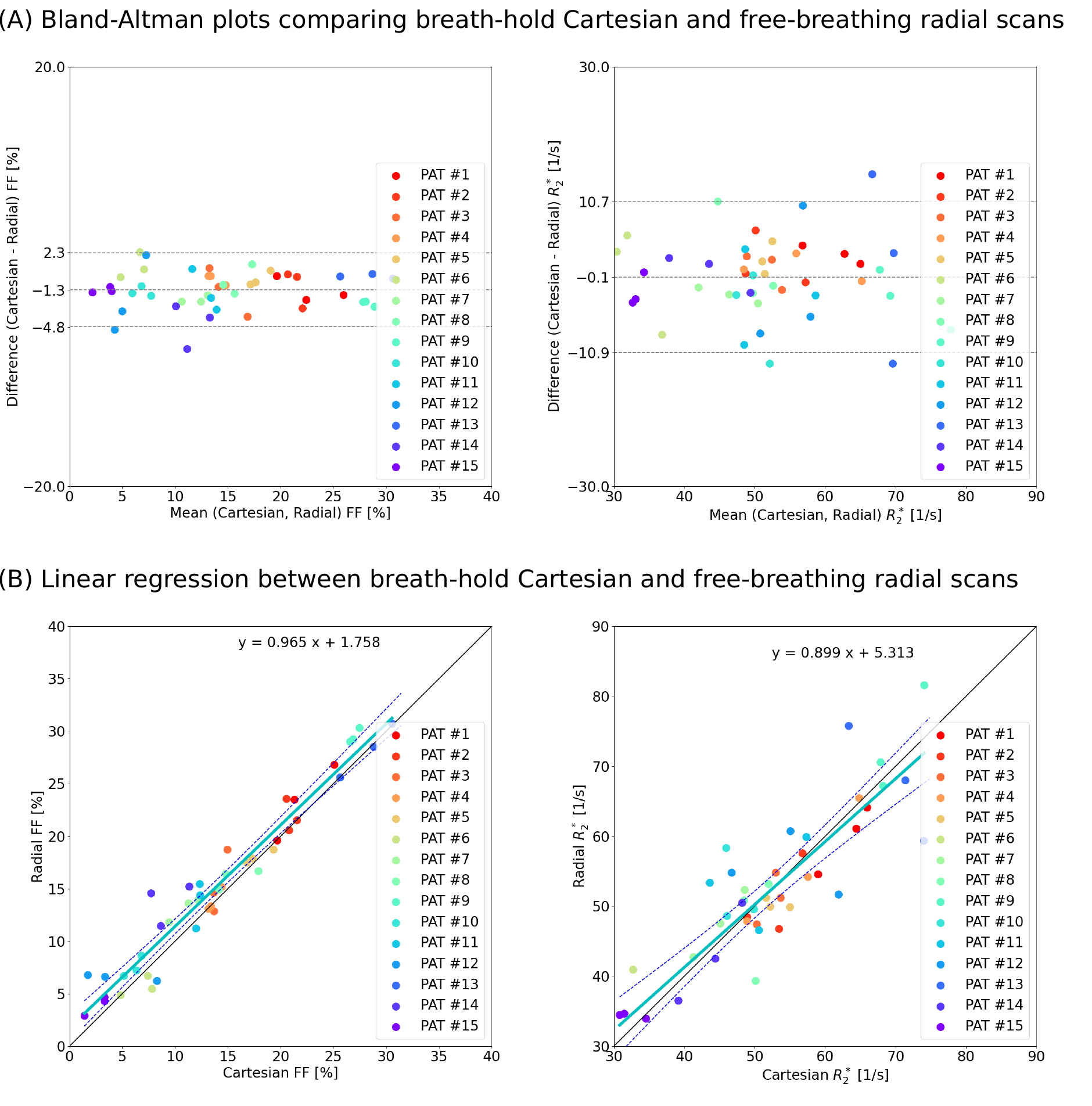}
	\caption{Quantitative analysis of reconstructed FF and $R_2^*$ maps 
		for all subjects comparing the reference breath-hold Cartesian scan 
		and the proposed free-breathing radial scan, respectively.}
	\label{FIG:LINREG}
\end{figure*}

\begin{table}[h!]
	\caption{Summary of FF values for all patients from the reference Cartesian and the proposed radial acquisition.}
	\label{TAB:SubjectSTAT_FF}
	\setlength{\tabcolsep}{3pt}
	\begin{tabular}{m{0.15\columnwidth} m{0.25\columnwidth} m{0.25\columnwidth} m{0.25\columnwidth}}
		\toprule
		& Cartesian & Radial (2:47) & Radial (1:24) \\
		\hline
		\multirow[t]{3}{*}{Patient \#1}   
		& $25.07 \pm 2.55$ & $26.81 \pm 3.62$ & $ 25.83 \pm 4.28$ \\ 
		& $19.66 \pm 1.92$ & $19.62 \pm 3.73$ & $ 20.47 \pm 3.67$ \\
		& $21.28 \pm 1.79$ & $23.50 \pm 2.32$ & $ 24.54 \pm 2.39$ \\
		\hline
		\multirow[t]{3}{*}{Patient \#2}   
		& $21.52 \pm 2.17$ & $21.53 \pm 2.34$ & $ 18.56 \pm 4.02 $ \\ 
		& $20.79 \pm 1.81$ & $20.56 \pm 3.08$ & $ 18.43 \pm 4.05 $ \\
		& $20.56 \pm 1.24$ & $23.58 \pm 1.33$ & $ 22.98 \pm 1.69 $ \\
		\hline
		\multirow[t]{3}{*}{Patient \#3}   
		& $13.66 \pm 2.31$ & $12.85 \pm 2.79$ & $15.36 \pm 3.88$ \\ 
		& $13.64 \pm 3.15$ & $14.61 \pm 3.42$ & $16.44 \pm 3.65$ \\
		& $14.95 \pm 1.76$ & $18.74 \pm 3.52$ & $19.89 \pm 4.04$ \\
		\hline
		\multirow[t]{3}{*}{Patient \#4}   
		& $13.18 \pm 3.55$ & $13.12 \pm 5.28$ & $13.64 \pm 5.97$ \\ 
		& $13.41 \pm 3.11$ & $13.34 \pm 2.91$ & $13.92 \pm 3.52$ \\
		& $14.37 \pm 3.04$ & $15.17 \pm 1.99$ & $15.45 \pm 2.27$ \\
		\hline
		\multirow[t]{3}{*}{Patient \#5}   
		& $16.77 \pm 1.35$ & $17.51 \pm 0.99$ & $13.31 \pm 1.66$ \\ 
		& $19.33 \pm 1.22$ & $18.74 \pm 1.29$ & $16.86 \pm 1.69$ \\
		& $17.34 \pm 2.05$ & $17.89 \pm 1.79$ & $17.04 \pm 1.87$ \\
		\hline
		\multirow[t]{3}{*}{Patient \#6}   
		& $ 7.82 \pm 2.90$ & $ 5.47 \pm 2.54$ & $7.98 \pm 3.73$ \\ 
		& $ 4.82 \pm 2.03$ & $ 4.87 \pm 2.12$ & $6.87 \pm 2.58$ \\
		& $ 7.39 \pm 1.23$ & $ 6.70 \pm 1.50$ & $8.52 \pm 2.45$ \\
		\hline
		\multirow[t]{3}{*}{Patient \#7}   
		& $ 9.44 \pm 1.28$ & $11.83 \pm 1.85$ & $ 9.34 \pm 3.12$ \\ 
		& $11.24 \pm 1.27$ & $13.63 \pm 1.70$ & $13.62 \pm 2.90$ \\
		& $12.18 \pm 1.59$ & $13.96 \pm 2.26$ & $13.36 \pm 3.22$ \\
		\hline
		\multirow[t]{3}{*}{Patient \#8}   
		& $17.87 \pm 3.10$ & $16.71 \pm 2.02$ & $16.45 \pm 3.71$ \\ 
		& $14.16 \pm 1.35$ & $14.94 \pm 3.71$ & $14.76 \pm 4.01$ \\
		& $14.80 \pm 1.40$ & $16.42 \pm 1.63$ & $16.79 \pm 2.89$ \\
		\hline
		\multirow[t]{3}{*}{Patient \#9}   
		& $26.60 \pm 0.76$ & $29.02 \pm 1.66$ & $28.71 \pm 2.19$ \\ 
		& $26.85 \pm 0.98$ & $29.24 \pm 2.21$ & $28.95 \pm 2.85$ \\
		& $27.45 \pm 1.05$ & $30.31 \pm 1.53$ & $30.08 \pm 2.34$ \\
		\hline
		\multirow[t]{3}{*}{Patient \#10}
		& $ 6.82 \pm 1.30$ & $ 8.65 \pm 2.71$ & $ 7.55 \pm 4.20$ \\ 
		& $ 6.33 \pm 0.92$ & $ 7.25 \pm 1.82$ & $ 6.41 \pm 3.12$ \\
		& $ 5.14 \pm 1.49$ & $ 6.73 \pm 2.00$ & $ 7.49 \pm 3.14$ \\
		\hline
		\multirow[t]{3}{*}{Patient \#11}
		& $12.38 \pm 3.38$ & $14.39 \pm 3.14$ & $15.34 \pm 4.05$ \\ 
		& $11.97 \pm 2.53$ & $11.24 \pm 4.08$ & $12.59 \pm 4.94$ \\
		& $12.33 \pm 1.30$ & $15.48 \pm 2.23$ & $13.83 \pm 4.02$ \\
		\hline
		\multirow[t]{3}{*}{Patient \#12}
		& $ 1.74 \pm 2.39$ & $ 6.81 \pm 3.68$ & $10.25 \pm 5.33$ \\ 
		& $ 3.36 \pm 3.51$ & $ 6.64 \pm 3.02$ & $ 8.05 \pm 3.87$ \\
		& $ 8.30 \pm 5.51$ & $ 6.23 \pm 2.63$ & $ 7.65 \pm 3.65$ \\
		\hline
		\multirow[t]{3}{*}{Patient \#13}
		& $25.62 \pm 1.68$ & $25.60 \pm 2.89$ & $28.64 \pm 2.86$ \\ 
		& $28.80 \pm 1.35$ & $28.55 \pm 2.19$ & $29.15 \pm 2.76$ \\
		& $30.55 \pm 1.78$ & $30.74 \pm 1.61$ & $31.25 \pm 3.41$ \\
		\hline
		\multirow[t]{3}{*}{Patient \#14}
		& $ 8.65 \pm 2.87$ & $11.48 \pm 2.01$ & $11.82 \pm 3.09$ \\ 
		& $ 7.71 \pm 2.44$ & $14.59 \pm 1.73$ & $15.27 \pm 2.20$ \\
		& $11.32 \pm 1.81$ & $15.21 \pm 1.46$ & $15.07 \pm 2.06$ \\
		\hline
		\multirow[t]{3}{*}{Patient \#15}
		& $ 1.41 \pm 1.04$ & $ 2.90 \pm 1.53$ & $ 8.23 \pm 2.94$ \\ 
		& $ 3.32 \pm 2.19$ & $ 4.31 \pm 1.95$ & $ 6.01 \pm 2.90$ \\
		& $ 3.30 \pm 1.64$ & $ 4.67 \pm 2.04$ & $ 4.96 \pm 2.41$ \\
		\bottomrule
	\end{tabular}
\end{table}

\begin{table}[h!]
	\caption{Summary of $R_2^*$ values for all patients from the reference Cartesian and the proposed radial acquisition.}
	\label{TAB:SubjectSTAT_R2S}
	\setlength{\tabcolsep}{3pt}
	\begin{tabular}{m{0.15\columnwidth} m{0.25\columnwidth} m{0.25\columnwidth} m{0.25\columnwidth}}
		\toprule
		& Cartesian & Radial (2:47) & Radial (1:24) \\
		\hline
		\multirow[t]{3}{*}{Patient \#1}   
		& $65.93 \pm 9.69$ & $64.08 \pm 11.38$ & $56.72 \pm 14.78$ \\ 
		& $59.00 \pm 7.67$ & $54.56 \pm  9.76$ & $54.21 \pm 13.88$ \\
		& $64.37 \pm 8.05$ & $61.13 \pm  7.22$ & $70.36 \pm  8.63$ \\
		\hline
		\multirow[t]{3}{*}{Patient \#2}   
		& $56.76 \pm 4.63$ & $57.58 \pm 13.52$ & $44.44 \pm 22.44$ \\ 
		& $53.43 \pm 6.51$ & $46.80 \pm  4.88$ & $48.66 \pm 5.86$ \\
		& $48.90 \pm 6.77$ & $48.46 \pm  4.23$ & $62.61 \pm 5.87$ \\
		\hline
		\multirow[t]{3}{*}{Patient \#3}   
		& $50.30 \pm  9.98$ & $47.42 \pm  9.91$ & $43.57 \pm 14.45$ \\ 
		& $53.66 \pm 11.92$ & $51.20 \pm  8.97$ & $52.30 \pm 11.97$ \\
		& $52.94 \pm  9.15$ & $54.81 \pm 10.26$ & $60.57 \pm 11.76$ \\
		\hline
		\multirow[t]{3}{*}{Patient \#4}   
		& $57.52 \pm 13.85$ & $54.21 \pm 13.28$ & $59.46 \pm 14.17$ \\ 
		& $48.93 \pm 17.38$ & $47.89 \pm 6.67$ & $47.05 \pm 7.71$ \\
		& $64.83 \pm  8.66$ & $65.47 \pm 5.54$ & $60.13 \pm 6.56$ \\
		\hline
		\multirow[t]{3}{*}{Patient \#5}   
		& $52.17 \pm 6.65$ & $49.96 \pm 4.98$ & $48.01 \pm 4.81$ \\ 
		& $51.61 \pm 6.70$ & $51.22 \pm 2.96$ & $52.09 \pm 3.83$ \\
		& $55.00 \pm 6.23$ & $49.92 \pm 3.57$ & $52.25 \pm 5.25$ \\
		\hline
		\multirow[t]{3}{*}{Patient \#6}   
		& $32.71 \pm 8.34$ & $40.98 \pm 8.37$ & $39.14 \pm 11.56$ \\ 
		& $32.20 \pm 5.82$ & $28.62 \pm 3.38$ & $27.12 \pm 6.21$ \\
		& $34.83 \pm 5.85$ & $28.93 \pm 2.94$ & $25.97 \pm 3.92$ \\
		\hline
		\multirow[t]{3}{*}{Patient \#7}   
		& $41.27 \pm 6.39$ & $42.79 \pm 5.27$ & $42.40 \pm 10.18$ \\ 
		& $48.56 \pm 4.66$ & $52.34 \pm 3.92$ & $46.31 \pm 7.66$ \\
		& $45.07 \pm 4.13$ & $47.59 \pm 6.75$ & $50.20 \pm 7.36$ \\
		\hline
		\multirow[t]{3}{*}{Patient \#8}   
		& $48.50 \pm 10.91$ & $50.84 \pm 6.26$ & $51.40 \pm 8.10$ \\ 
		& $50.11 \pm  6.90$ & $39.33 \pm 13.85$ & $43.56 \pm 15.84$ \\
		& $51.94 \pm  4.19$ & $53.21 \pm 5.36$ & $50.50 \pm 7.45$ \\
		\hline
		\multirow[t]{3}{*}{Patient \#9}   
		& $68.23 \pm 3.42$ & $67.24 \pm 10.13$ & $72.99 \pm 14.40$ \\ 
		& $67.82 \pm 4.67$ & $70.57 \pm 8.16$ & $74.60 \pm 10.89$ \\
		& $74.05 \pm 4.03$ & $81.62 \pm 6.64$ & $77.14 \pm 11.25$ \\
		\hline
		\multirow[t]{3}{*}{Patient \#10}
		& $49.84 \pm 6.23$ & $49.62 \pm 3.38$ & $49.21 \pm 7.14$ \\ 
		& $46.05 \pm 3.01$ & $48.66 \pm 2.99$ & $50.22 \pm 9.66$ \\
		& $45.90 \pm 4.85$ & $58.33 \pm 4.70$ & $70.90 \pm 8.42$ \\
		\hline
		\multirow[t]{3}{*}{Patient \#11}
		& $43.59 \pm 13.09$ & $53.36 \pm 7.96$ & $47.71 \pm 12.44$ \\ 
		& $50.56 \pm  6.90$ & $46.61 \pm 7.59$ & $49.93 \pm  9.97$ \\
		& $57.28 \pm  6.31$ & $59.93 \pm 3.69$ & $61.10 \pm  6.77$ \\
		\hline
		\multirow[t]{3}{*}{Patient \#12}
		& $46.69 \pm 30.69$ & $54.82 \pm 11.62$ & $53.38 \pm 15.59$ \\ 
		& $55.03 \pm 15.29$ & $60.74 \pm  7.26$ & $50.06 \pm 14.66$ \\
		& $61.89 \pm 22.54$ & $51.72 \pm  8.29$ & $46.04 \pm 10.79$ \\
		\hline
		\multirow[t]{3}{*}{Patient \#13}
		& $63.34 \pm 9.53$ & $75.77 \pm 10.40$ & $64.83 \pm 13.06$ \\ 
		& $74.00 \pm 4.67$ & $59.35 \pm 10.18$ & $56.07 \pm 19.94$ \\
		& $71.38 \pm 9.57$ & $68.00 \pm  6.87$ & $66.23 \pm 10.89$ \\
		\hline
		\multirow[t]{3}{*}{Patient \#14}
		& $39.16 \pm 12.64$ & $36.50 \pm 3.20$ & $33.93 \pm 5.64$ \\ 
		& $44.40 \pm  6.55$ & $42.55 \pm 3.33$ & $47.10 \pm 5.59$ \\
		& $48.21 \pm  3.82$ & $50.52 \pm 3.96$ & $58.19 \pm 6.96$ \\
		\hline
		\multirow[t]{3}{*}{Patient \#15}
		& $31.44 \pm 2.53$ & $34.66 \pm 5.33$ & $28.00 \pm 6.30$ \\ 
		& $34.58 \pm 7.58$ & $33.96 \pm 4.72$ & $31.77 \pm 5.86$ \\
		& $30.83 \pm 3.96$ & $34.52 \pm 3.90$ & $30.13 \pm 6.54$ \\
		\bottomrule
	\end{tabular}
\end{table}

\Cref{FIG:LINREG} depicts the Bland-Altman as well as scatter plots 
comparing the reference breath-hold Cartesian scan and 
the proposed free-breathing radial scan. 
For this quantitative analysis, three regions of interest (ROI) 
were selected for every subject and the mean value of every ROI 
was used for plotting. 
In the Bland-Altman plots, the x and y axis represents 
the mean and difference values between these two scans, 
respectively.
The central dotted line along the x-axis shows the mean bias 
between these two scans and is computed by averaging the difference values.
The upper and lower dotted lines show the limits of agreement 
(mean bias~$\pm$~1.96~$\times$~standard deviation of difference). 
In the scatter plots, the x and y axis represents the mean values of 
the Cartesian and the radial scan, respectively. 
In addition, linear regression was performed. 
The slope of the fitted curve for both FF and $R_2^*$ values is close to 1, 
indicating a good match between the Cartesian and the radial measurements.

\cref{TAB:SubjectSTAT_FF,TAB:SubjectSTAT_R2S} further summarize 
both the mean and the standard deviation values of every ROI and every patient. 
Overall, the patients in this work cover wide ranges of FF and $R_2^*$ values, 
and hence are representative for the validation of the proposed method. 
Moreover, the quantitative analysis results reveal two important clinical indications.

\begin{figure}
	\centering
	\includegraphics[width=\columnwidth]{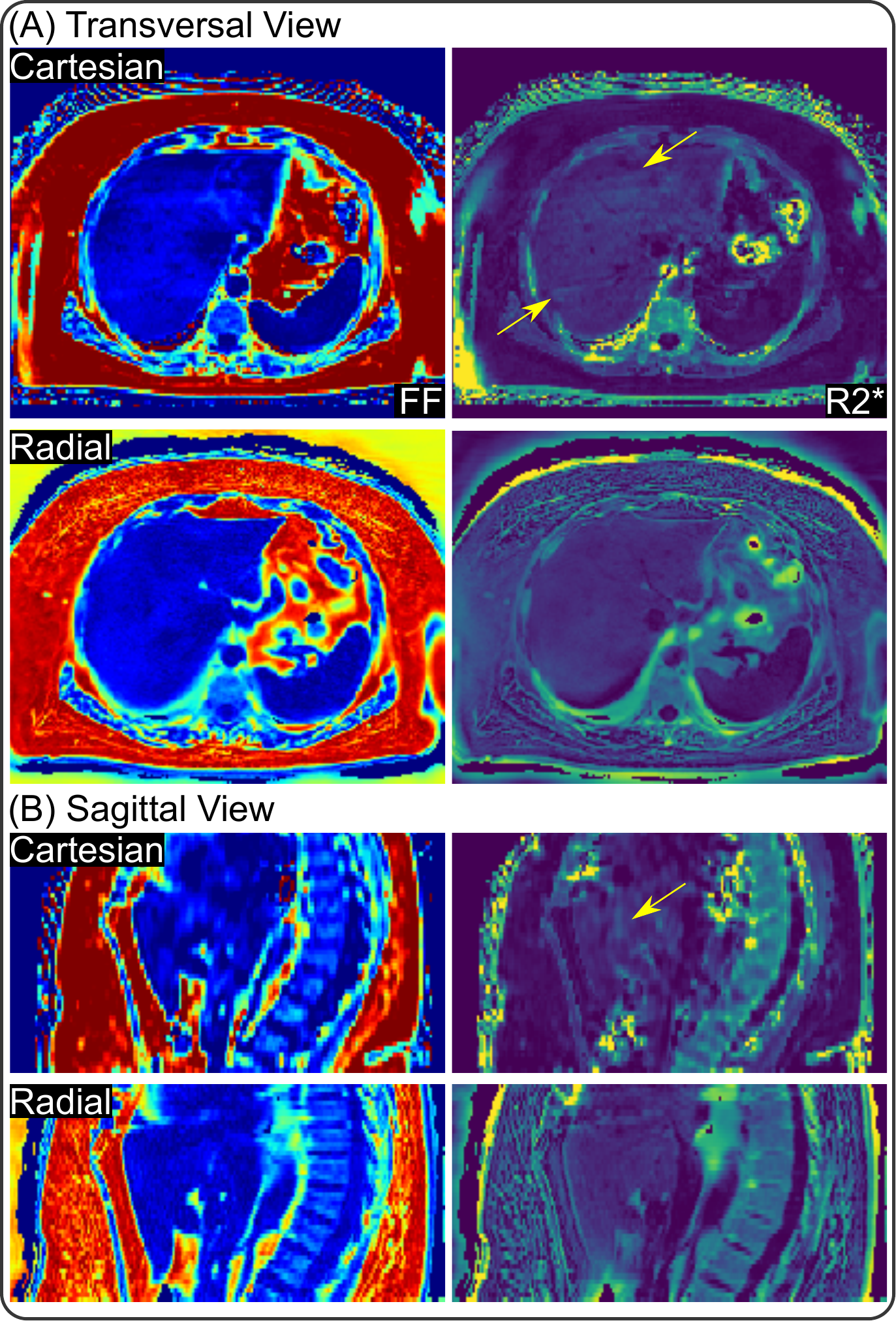}
	\caption{(A) Transversal view and (B) sagittal view 
		of reconstructed FF and $R_2^*$ maps in Patient \#6 
		comparing the reference breath-hold Cartesian scan and 
		the proposed radial scan, respectively. 
		This patient shows definite obesity symptom 
		(see \cref{TAB:Subject}), but has no fatty liver. 
		In fact, the FF values of this patient is the lowest 
		among all subjects (see also \cref{FIG:LINREG}). 
		Similar to the results of Patient \#1 in \cref{FIG:REGU}, 
		the Cartesian scan suffers from fold-in artifacts in the $R_2^*$ maps 
		(yellow arrows), while the proposed radial scan shows consistent $R_2^*$ values.}
	\label{FIG:LPDFF}
\end{figure}

First, obese patients may not necessarily have fatty liver disease. 
Patient \#6 is diagnosed with obesity, 
and shows the highest BMI in \cref{TAB:Subject}. 
For this patient, however, ultrasound is not able to penetrate through 
the thick subcutaneous fat layer to assess the liver fat content. 
As shown in \cref{FIG:LPDFF}, 
both breath-hold Cartesian and free-breathing radial scans are able to 
provide quantitative FF and $R_2^*$ maps. The liver regions exhibit 
relatively low FF values. 
This patient shows the lowest FF and $R_2^*$ values 
among all subjects in this study (see \cref{FIG:LINREG}). 
On the other hand, the $R_2^*$ maps 
in both the transversal and sagittal views 
again suffer from hyper intensities in the Cartesian scan 
(as pointed by yellow arrows), 
while our proposed radial scan shows more homogeneous $R_2^*$ maps.

\begin{figure*}
	\centering
	\includegraphics[width=0.70\textwidth]{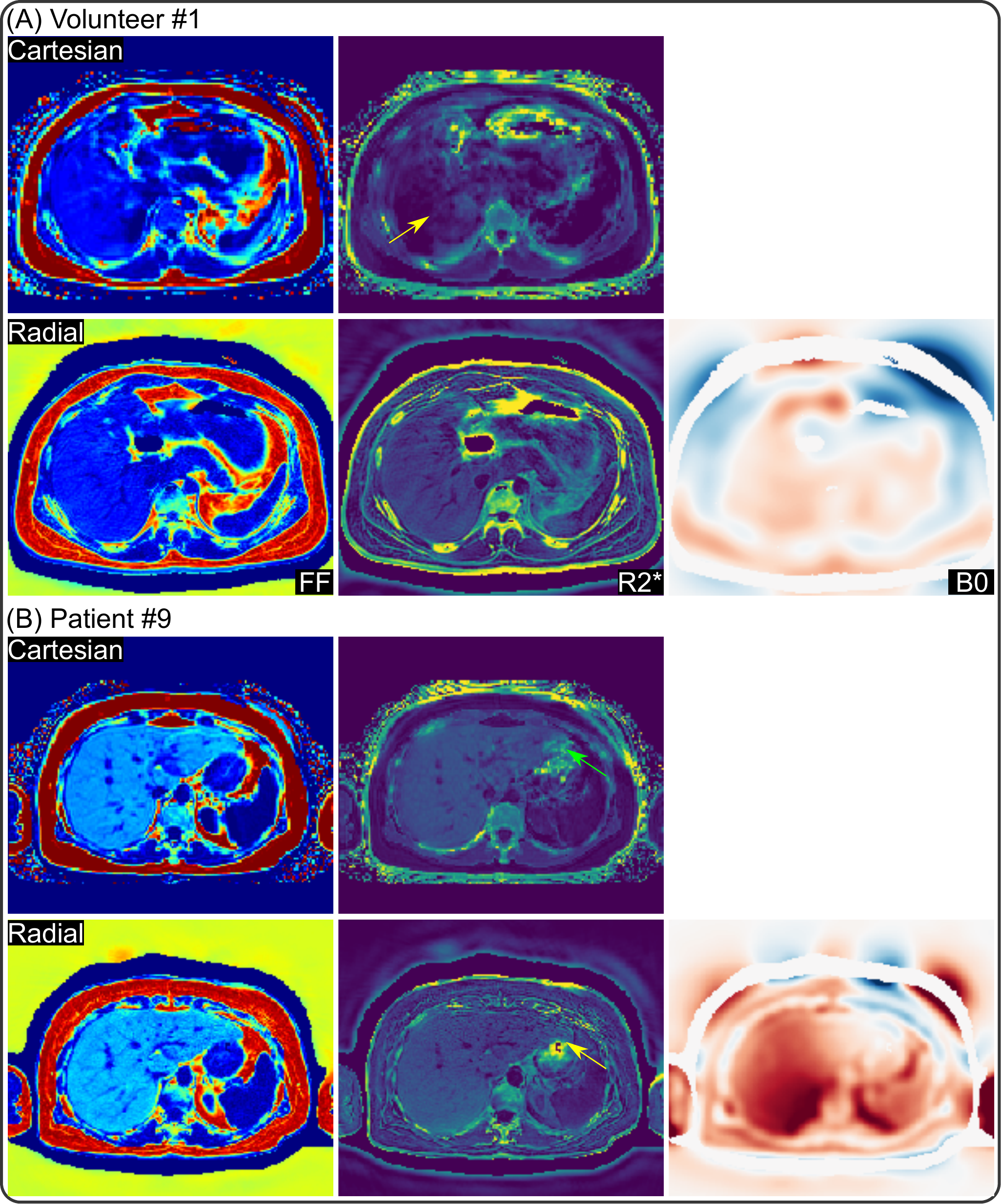}
	\caption{(A) Volunteer \#1 and (B) Patient \#9 
		with elevated FF values, but no clear symptom of obesity.
		In particular, Patient \#9 was diagnosed with hepatic steatosis 
		by the standard liver biopsy and ultrasound.}
	\label{FIG:HPDFF}
\end{figure*}

Second, in our study, patients with hepatic steatosis may not necessarily be obese. 
\Cref{FIG:HPDFF} displays one volunteer and one patient with elevated FF values from 
both the reference breath-hold Cartesian and the free-breathing radial 
scans. Patient \#9 has hepatic steatosis as confirmed via 
liver biopsy and ultrasound diagnosis. 
However, neither of them shows the problem of obesity (see \cref{TAB:Subject}). 
Through \cref{FIG:LINREG} and \cref{FIG:HPDFF}, 
we also observe that hepatic steatosis is more likely to occur in elderly patients. 
Although fatty liver is also captured in the scan of Volunteer \#1, 
this is rather a rare case among all young volunteers (results not shown here). 
On the other hand, in the left corner of the liver in Patient \#9 
(see the green arrows), 
$R_2^*$ values from the Cartesian scan again are higher, 
possibly due to the fast $B_0$ change in this air-tissue interface, 
created by the empty region within the stomach.

\begin{figure*}
	\centering
	\includegraphics[width=\textwidth]{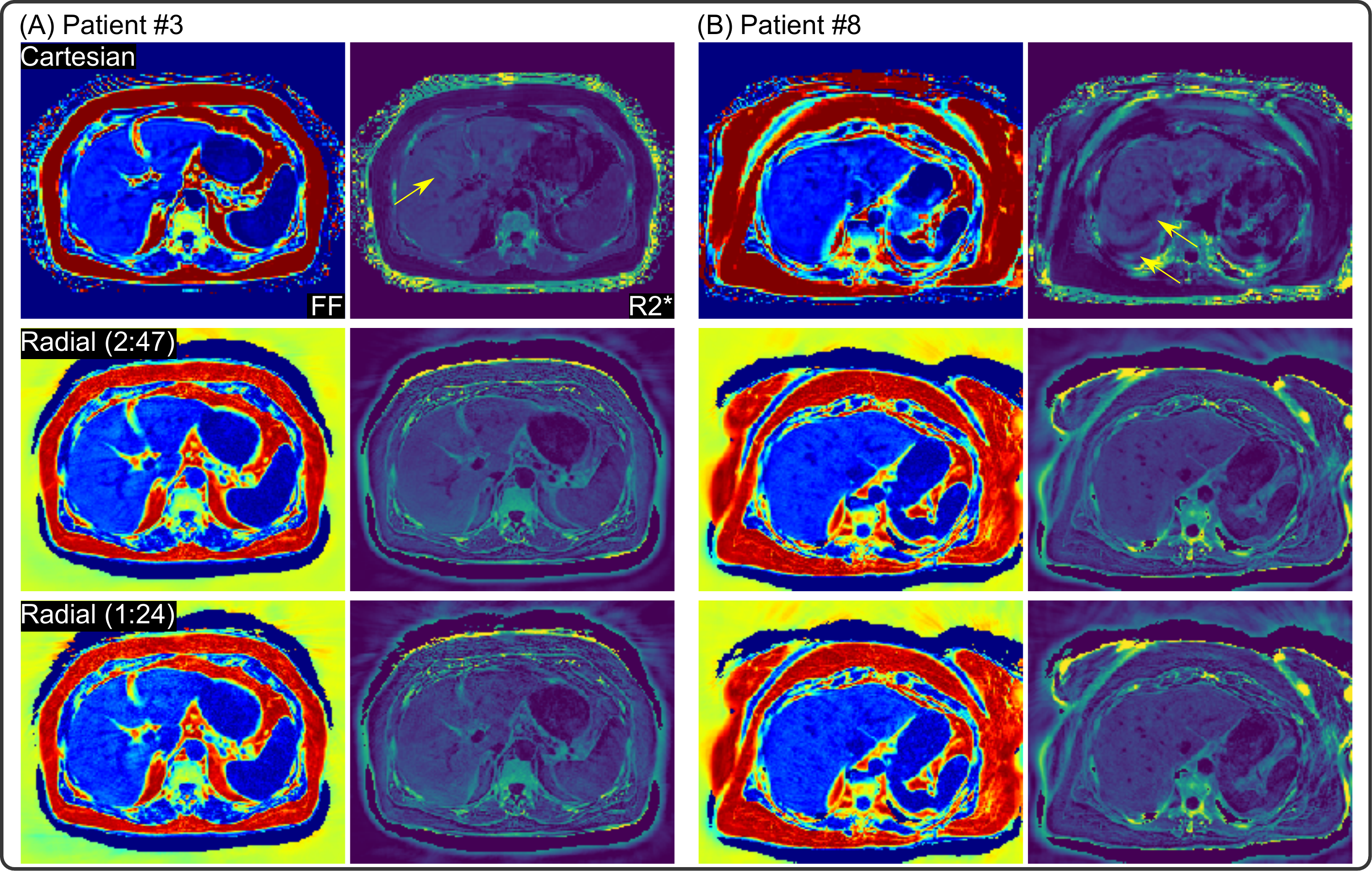}
	\caption{Comparison of FF and $R_2^*$ maps for Patients (A) \#3 and (B) \#8 
		among the reference Cartesian scan, 
		the free-breathing radial 2:47~min scan, 
		and the free-breathing radial scan with retrospective two-fold undersampling,
		corresponding to a scan time of 1:24~\si{\minute}.
		Note the ripple-like artifact in the $R_2^*$ map from the Cartesian scans 
		due to incomplete breath hold (yellow arrows in the 1st row).}
	\label{FIG:R2STAR}
\end{figure*}

Noteworthy, the reference Cartesian scan requires subjects to hold their breath. 
When subjects fail to perform breath holding, 
it may lead to image artifacts in the reconstruction. 
\Cref{FIG:HPDFF,FIG:R2STAR} (see yellow arrows) are such examples. 
Residual respiratory motion during the scan 
causes fast $B_0$ field drift \cite{tan_2019_mobawf} 
as well as signal variation. 
This situation makes the fitting of $B_0$ and $R_2^*$ difficult. 
In this example, it results in the blurring artifact for Patient \#3 
and ripple-like artifact for Patient \#8 (see yellow arrows) 
and unreliable $R_2^*$ values. 
However, the proposed free-breathing radial scan is free of such artifacts.

\begin{figure*}
	\centering
	\includegraphics[width=0.75\textwidth]{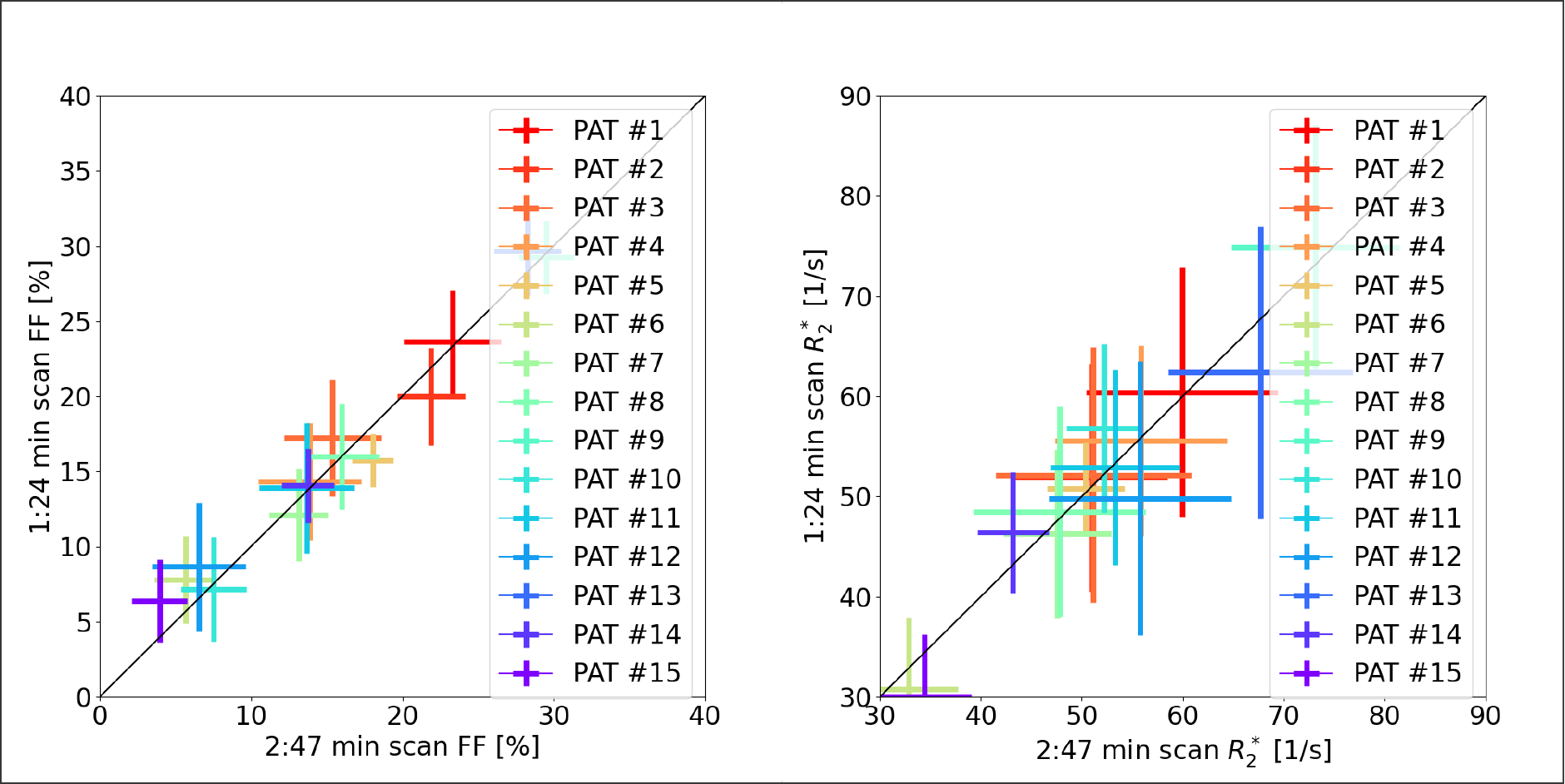}
	\caption{Quantitative analysis of the reconstructed FF and $R_2^*$ maps 
		from radial scans:~actual acquisition (2:47~\si{\minute}) and 
		retrospectively undersampled acquisition (1:24~\si{\minute}), 
		respectively.}
	\label{FIG:USMP}
\end{figure*}

Furthermore, we investigated reducing the acquired radial data by one-half, 
to explore the feasibility of the proposed motion-resolved model-based reconstruction 
on retrospectively undersampled data. 
As shown in \cref{FIG:R2STAR,FIG:USMP} as well as in the right column of 
\cref{TAB:SubjectSTAT_FF,TAB:SubjectSTAT_R2S}, retrospective undersampling 
increases standard deviation in the reconstructed FF and $R_2^*$ maps, 
but the mean values agree well with the reconstruction on the 2:47~min scan (the center column).

\begin{figure}
	\centering
	\includegraphics[width=\columnwidth]{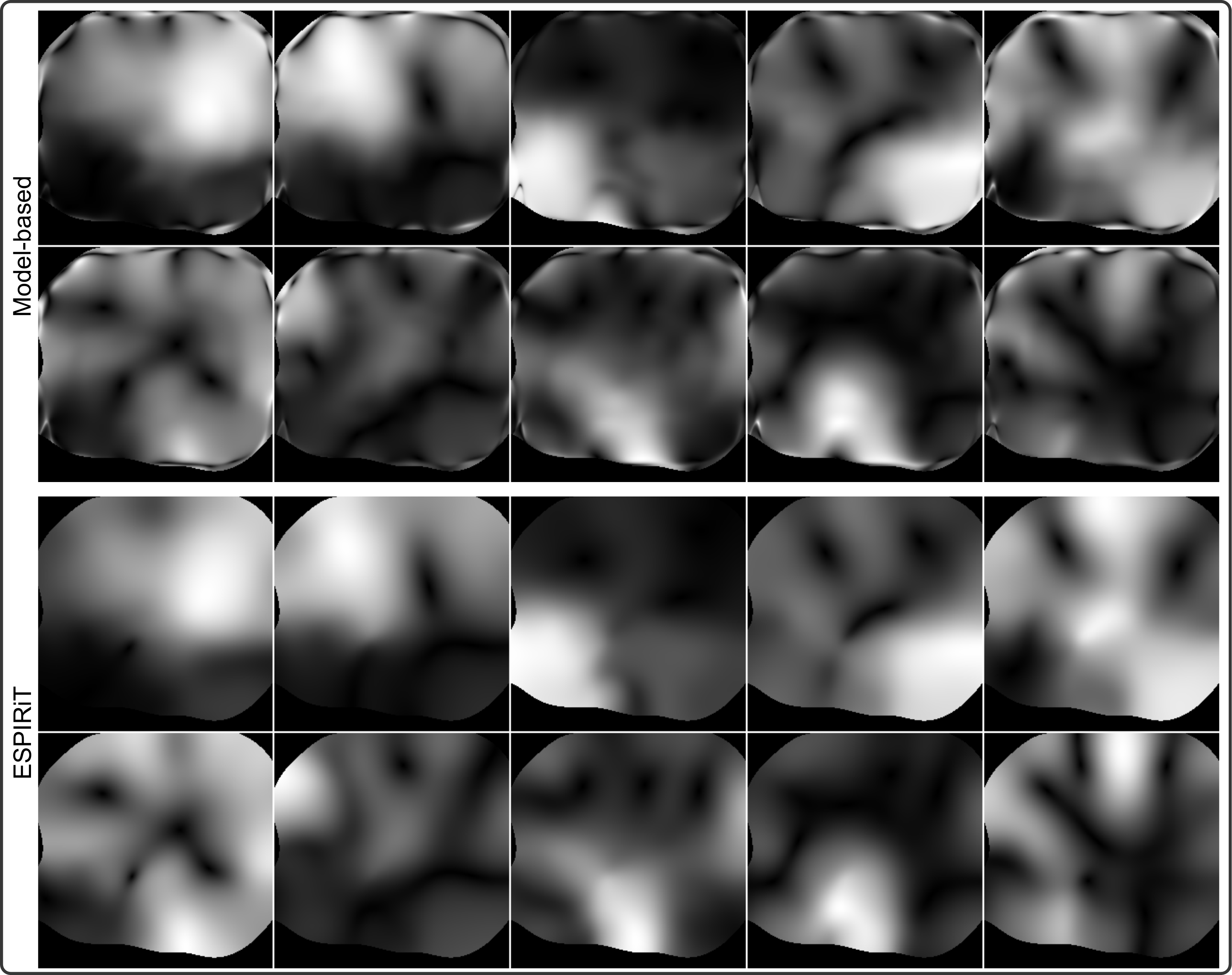}
	\caption{Magnitude of coil sensitivity maps for Patient \#1 from 
		(top) the proposed model-based method and (bottom) the ESPIRiT method.}
	\label{FIG:SENS}
\end{figure}

This work uses joint estimation of all parameter maps by solving a
nonlinear inverse problem. This includes simultaneous estimation of the
coil sensitivity maps.
\cref{FIG:SENS} displays the ten coil sensitivity maps 
of the first respiratory bin in comparison to sensitivities estimated
with ESPIRiT \cite{uecker_2014_espirit}.
Since coil sensitivity maps from ESPIRiT are inherently normalized, 
normalization is also applied to the coil sensitivity maps from the
model based reconstruction. The coil sensitivity maps from both
methods are very similar inside the region with signal.

\section{Discussion}

This work introduced a stack-of-radial multi-echo radial FLASH sequence 
with partial Fourier readouts (i.e.~asymmetric echo) 
for volumetric free-breathing liver acquisition. 
Moreover, this work introduced a regularized model-based reconstruction 
in BART to jointly estimate liver water, fat, $R_2^*$, 
$B_0$ field inhomogeneity maps, and coil sensitivity maps 
directly from acquired $k$-space data. 
This nonlinear inverse problem was solved by IRGNM with ADMM, 
allowing for generalized regularization terms in nonlinear reconstruction. 
To validate the proposed MRI sequence and reconstruction, 
a pilot study was conducted comprising young volunteers and 
patients diagnosed with obesity, diabetes, or hepatic steatosis. 
The reconstructed FF and $R_2^*$ maps were further compared with 
the reference breath-hold Cartesian scan. 
We found good agreement between the proposed free-breathing radial scan 
and the breath-hold Cartesian scan.

The $R_2^*$ maps of Patient \#1 from the proposed method are visually more blurred 
compared to the reference Cartesian scan, 
while the volunteer $R_2^*$ map shows a sharp delineation of the liver and vessel borders. 
This may be due to the use of a larger FOV in the acquisition protocol for patients, 
which corresponds to higher undersampling. 
On the other hand, the larger body size of this patient 
may result in poorer receiver signals in the middle.

In model-based reconstructions, parameter maps are directly reconstructed 
from acquired $k$-space data. 
Therefore, regularizations can be directly imposed on parameter maps. 
An alternative approach is to reconstruct all echo images 
and then perform image-space parameter fitting 
\cite{armstrong_2018_fat,zhong_2020_r2sclinic,zhong_2020_resr2s,wang_2022_mt-me,starekova_2022_fb-fat}.
In this case, advanced regularizations (e.g.~$\ell^1$-Wavelet and low rankness) 
can be employed along the echo dimension. 
However, this study mainly focused on the comparison 
with the reference breath-hold Cartesian scan. 
In addition, this study also compared against another model-based reconstruction approach 
using temporal total variation regularization on water, fat and $R_2^*$ maps, 
whereas $B_0$ and coil sensitivity maps were calibrated and kept constant  \cite{schneider_2020_mobawfr2s}.

This work initialized the $B_0$ field inhomogeneity map for every respiratory bin 
by the $B_0$ estimate from a 3-echo model-based reconstruction. 
As only three echoes were used for the reconstruction, 
this initialization procedure is relatively fast. 
Such $B_0$ initialization assures the convergence of large-scale model-based reconstructions 
on seven echoes and four respiratory bins. 
Recently, Zhang et al.~proposed the solution interval method for nonlinear inverse problems 
\cite{zhang_2021_nlls_init}. 
However, abundant computation power is required to 
run the nonlinear least square fitting multiple times 
for the construction of solution intervals for every unknown.
In general, nonlinear inverse problems consisting of non-convex phase variation 
is not trivial to be solved. 
Similar problems exist in multi-shot diffusion MRI as well. 
Recently, Hu et al.~\cite{hu_2020_spa_llr} proposed to 
jointly estimate diffusion-weighted magnitude and phase images 
to resolve the shot-to-shot phase variation. 
Similar to our work, this approach solves a non-linear non-convex inverse problem, 
but still requires good phase initialization.

There are several aspects of this study that can be further improved. 
No patient participated in this study 
showed symptoms of iron overload (i.e.~elevated $R_2^*$ values). 
To the best of our knowledge, iron overload is more likely to be seen 
in patients who require regular blood transfusions. 
These patients, however, are rather rare in obesity/diabetes clinics. 
Although lacking such patients, we showed the capability of our technique in 
quantifying iron overload in the phantom, as seen in \cref{FIG:PHA}, 
where Tube 3 results in the highest $R_2^*$ value due to iron. 

Joint estimation of all physical parameter maps 
and coil sensitivity maps is a nonlinear nonconvex inverse problem. 
Especially when including the $B_0$ field inhomogeneity map as one unknown, 
the reconstruction is sensitive to the initial guess and scaling of unknowns. 
Our implementation initialized the $B_0$ map 
with the previously proposed three-echo reconstruction \cite{tan_2019_mobawf}, 
which indeed fostered convergence. 
Second, scaling of the fat, $R_2^*$, and $B_0$ maps was 
empirically determined. Therefore, more intelligent $B_0$ estimation 
and automatic scaling of unknowns \cite{tan_2017_scalemobaflow} 
would be logical directions. 

Compared with the reference breath-hold Cartesian scan, 
our proposed radial sampling did not explore 
$k_z$ unaligned undersampling strategies \cite{breuer_2005_caipi}, 
which can further accelerate data acquisition.

While this work focused on liver fat and $R_2^*$ mapping, 
the presented multi-echo radial sequence in principle 
is also applicable to brain imaging, which, 
would become rather interesting in combination with 
the above-mentioned $k_z$ undersampling. 
Alternatively, one can also explore the possibility of 
multiple parametric mapping 
via extending the basic multi-echo radial acquisition to 
variable flip angles or magnetization preparation.

This work employed the recently proposed SSA-FARY technique 
for self-gating, where the determination of spokes as respiratory bins 
was based on polar angles in the phase portrait plot. 
This procedure did not take into account the radius of every spoke 
in phase portrait, which might cause problems in the case of irregular breathing. 
Therefore, it might make the self-gating technique more general considering the radius. Alternatively, it would be interesting to use 
pilot tone \cite{solomon_2021_pilot_tone} for prospective respiratory binning 
during data acquisition. 

The total reconstruction time for the whole liver took about \SI{4}{\hour} 
on the Tesla V100 GPU. For clinical translation, 
further acceleration could be achieved via 
for example multiple GPU parallelization. 
This includes two aspects. 
First, the proposed reconstruction algorithm 
can be parallelized with multiple GPUs 
to accelerate the iterative minimization procedure.
Second, currently slices are reconstructed in a sequential manner. 
This can be modified to distribute the slice reconstruction in parallel 
to further speed up the whole process.
These aspects, however, are outside the scope of this work, 
which mainly focused on the technical development 
of the free-breathing model-based reconstruction method.

\section{Conclusion}

This work introduced a free-breathing liver fat and $R_2^*$ quantification 
technique, comprising stack-of-radial multi-echo asymmetric-echo volumetric continuous acquisition 
and regularized non-linear model-based reconstruction. 
The generic model-based reconstruction framework in BART allows the flexible
use of generalized regularization terms and is integrable with different 
physical models. This work uses this framework for joint estimation of
time-resolved physical parameter maps (water, fat, $R_2^*$, and $B_0$) 
and coil sensitivity maps.
The proposed method is validated against a reference breath-hold Cartesian scan 
on healthy volunteers and patients.
This technique offers a non-invasive tool for quantitative liver assessment
during free breathing.

\begin{appendices}
	\crefalias{section}{appsec}
	
	\counterwithin{figure}{section}
	\counterwithin{table}{section}
	
	\section{}
	\label{SEC:APPENDIXA}
	
	\setcounter{equation}{0}
	\renewcommand{\theequation}{A.\arabic{equation}}
	
	\noindent To minimize the cost function in \cref{EQU:obj_func}, 
	IRGNM linearizes the nonlinear forward model via Taylor expansion in each Newton step, 
	thus the data-consistency term in \cref{EQU:obj_func} becomes
	\begin{equation} \label{EQU:lin_problem}
		\norm{DF(x_n) \text{d} x - [ y - F(x_n) ] }_2^2
	\end{equation}
	where the Jacobian $DF(x_n)$ denotes the derivative of the forward operator 
	concerning the $n$th-step estimate. 
	Given the initial guess as $x_0$ \cite{uecker_2008_nlinv}, 
	one can denote $x = x_{n+1} - x_0 = x_{n} + \text{d} x - x_0$. 
	As a result, \cref{EQU:lin_problem} becomes
	\begin{equation}
		\begin{aligned}
			&\norm{DF(x_n) (x + x_0 - x_n) - [ y - F(x_n) ] }_2^2 \\
			\Rightarrow &\norm{DF(x_n) x -[ DF(x_n) (x_n - x_0) +  y - F(x_n) ] }_2^2
		\end{aligned}
	\end{equation}
	whose minimum occurs when its derivative is set to $0$, and we obtain such a linear system equation,
	\begin{equation} \label{EQU:lsqr}
		A x = b
	\end{equation}
	for which we denote $A := DF^H (x_n) DF(x_n)$ and $b := DF^H (x_n) \big\{ DF(x_n) (x_n - x_0) +  y - F(x_n) \big\}$.
	
	With generalized $\ell1$ regularization, \cref{EQU:lsqr} can be written in the ADMM form,
	\begin{equation} \label{EQU:ADMM_LASSO}
		\begin{split}
			\textrm{minimize}\;\; &\norm{Ax - b}_2^2 + \alpha \norm{z}_1 \\
			\textrm{subject to}\;\; &Tx - z = 0
		\end{split}
	\end{equation}
	The updates can be derived,
	\begin{equation} \label{EQU:ADMM_LASSO_upd}
		\left\{\begin{matrix}
			\begin{aligned}
				x^{(k+1)} &:= (A^H A + \rho T^H T / 2) [A^H b + \rho T^H (z^{(k)} - \mu^{(k)})/2] \\ 
				z^{(k+1)} &:= \mathcal{T}_{\alpha/\rho} (T x^{(k+1)} + \mu^{(k)}) \\
				u^{(k+1)} &:= u^{(k)} + T x^{(k+1)} - z^{(k+1)}
			\end{aligned}
		\end{matrix}\right.
	\end{equation}
	The $x$ update is solved by the conjugate gradient method, 
	and the $z$ update is computed via soft thresholding ($\mathcal{T}_{\alpha/\rho}$), 
	where $\alpha$ is passed from IRGNM and iteratively reduced along Newton steps, 
	$\alpha = 1/D^{n-1}$ with $D > 1$ and $n$ the $n$the Newton iteration. 
	$\rho$ is known as the penalty parameter in ADMM. 
	
	\section{}
	\label{SEC:APPENDIXB}
	
	The iterative solution to \cref{EQU:lsqr} requires the computation of the Jacobian $DF(x)$ 
	and its corresponding adjoint $DF^H (x)$ operator, with the forward operator $F(x)$ denoted 
	in \cref{EQU:op_fwd}. 
	Note that the forward operator can be split into 
	two nonlinear operators:~the parallel imaging operator ($P\mathcal{F} M \mathcal{S}$) 
	and multi-echo signal model operator ($\mathcal{B}$). 
	Since the first one has already been implemented 
	in BART for parallel imaging as nonlinear inversion (NLINV) \cite{uecker_2008_nlinv}, 
	only the second operator is required to be implemented.
	Afterward, the two nonlinear operators can be chained together. 
	Therefore, only the operator $\mathcal{B}$ is explained in detail here.
	
	As denoted in \cref{EQU:meco_wfr2s}, the nonlinear operator $\mathcal{B}$ presents the mapping 
	from the parameter maps ($\text{W}, \text{F}, R_2^*, f_{B_0}$) 
	to the multi-echo images ($\rho_m$), thus 
	\begin{equation} \label{EQU:B_mapping}
		\mathcal{B} : \mathbb{C}^{N^2 \times N_p} \mapsto \mathbb{C}^{N^2 \times E} \; .
	\end{equation}
	Here, $N^2$ denotes the image size, $N_p$ the number of parameter maps 
	(4 in this case) and $E$ the number of echoes.
	Therefore, its Jacobian matrix $D \mathcal{B} \in \mathbb{C}^{N^2 \times E \times N_p}$. 
	Denote $\mathcal{B}_m$ as the operator output corresponding to the $m$th TE, 
	its corresponding Jacobian is
	\begin{multline*}
		D\mathcal{B}_m = 
		\begin{pmatrix}
			\frac{\partial \mathcal{B}_m}{\partial \text{W}} \\
			\frac{\partial \mathcal{B}_m}{\partial \text{F}} \\
			\frac{\partial \mathcal{B}_m}{\partial {R_2^*}} \\
			\frac{\partial \mathcal{B}_m}{\partial f_{B_0}}
		\end{pmatrix}^T \\ = 
		\begin{pmatrix}
			\begin{array}{r}
				e^{-{R_2^*} \text{TE}_m} \cdot e^{i2\pi f_{B_0} \text{TE}_m} \\
				z_m \cdot e^{-{R_2^*} \text{TE}_m} \cdot e^{i2\pi f_{B_0} \text{TE}_m} \\
				\text{F} \cdot z_m \cdot \Big( (-\text{TE}_m) e^{-{R_2^*} \text{TE}_m} \Big) \cdot e^{i2\pi f_{B_0} \text{TE}_m} \\
				\Big( \text{W} + \text{F} \cdot z_m \Big) \cdot e^{-{R_2^*} \text{TE}_m} \cdot (i2\pi \text{TE}_m) \cdot e^{i2\pi f_{B_0} \text{TE}_m}
			\end{array}
		\end{pmatrix}^T
	\end{multline*}
	The adjoint operator is then its complex conjugate transpose.
	
\end{appendices}

\section*{Acknowledgment}

Z.~T.~sincerely thanks Ms.~Sarina Tepan 
for help with the organization of patient study. 
Z.~T.~also thanks Drs.~Jens Frahm and Kai Tobias Block 
for various discussions.
Z.~T.~and M.~U.~thank DFG for the research grant 
TA 1473/2-1 and UE 189/4-1.

\bibliography{ref}

\begin{thebibliography}{51}
\providecommand{\natexlab}[1]{#1}
\providecommand{\url}[1]{\texttt{#1}}
\providecommand{\urlprefix}{}

\bibitem[{Caussy et~al.(2018)Caussy, Cyrielle and Reeder, Scott B. and Sirlin,
  Claude B. and Loomba, Rohit}]{caussy_2018_fat}
Caussy C, Reeder SB, Sirlin CB, Loomba R.
\newblock {Noninvasive, quantitative assessment of liver fat by MRI-PDFF as an
  endpoint in NASH trials}.
\newblock Hepatology 2018;68:763--772.

\bibitem[{Hu et~al.(2020)Hu, Houchun H. and Branca, Rosa Tamara and Hernando,
  Diego and Karampinos, Dimitrios C. and Machann, J\"{u}rgen and McKenzie,
  Charles A. and Wu, Holden H. and Yokoo, Takeshi and Velan, S.
  Sendhil}]{hu_2020_obesity}
Hu HH, Branca RT, Hernando D, Karampinos DC, Machann J, McKenzie CA, et~al.
\newblock {Magnetic resonance imaging of obesity and metabolic disorders:
  Summary from the 2019 ISMRM Workshop}.
\newblock Magn Reson Med 2020;83:1565--1576.

\bibitem[{Wood(2011)Wood, John C}]{wood_2011_iron}
Wood JC.
\newblock {Impact of iron assessment by MRI}.
\newblock Hematology 2011;2011:443--450.

\bibitem[{Hernando et~al.(2014)Hernando, Diego and Levin, Yakir S and Sirlin,
  Claude B and Reeder, Scott B}]{hernando_2014_iron}
Hernando D, Levin YS, Sirlin CB, Reeder SB.
\newblock {Quantification of liver iron with MRI: State of the art and
  remaining challenge}.
\newblock J Magn Reson Imaging 2014;40:1003--1021.

\bibitem[{Dixon(1984)Dixon, W Thomas}]{dixon_1984_wf}
Dixon WT.
\newblock {Simple proton spectroscopic imaging}.
\newblock Radiology 1984;153:189--194.

\bibitem[{Armstrong et~al.(2018)Armstrong, Tess and Dregely, Isabel and
  Stemmer, Alto and Han, Fei and Natsuaki, Yutaka and Sung, Kyunghyun and Wu,
  Holden H.}]{armstrong_2018_fat}
Armstrong T, Dregely I, Stemmer A, Han F, Natsuaki Y, Sung K, et~al.
\newblock {Free-breathing liver fat quantification using a multiecho 3D
  stack-of-radial techqniue}.
\newblock Magn Reson Med 2018;79:370--382.

\bibitem[{Zhong et~al.(2020{\natexlab{a}})Zhong, Xiaodong and Hu, Houchun H.
  and Armstrong, Tess and Li, Xinzhou and Lee, Yu-Hsiu and Tsao, Tsu-Chin and
  Nickel, Marcel D. and Kannengiesser, Stephan A.R. and Dale, Brian M. and
  Deshpande, Vibhas and Kiefer, Berthold and Wu, Holden
  H.}]{zhong_2020_r2sclinic}
Zhong X, Hu HH, Armstrong T, Li X, Lee YH, Tsao TC, et~al.
\newblock {Free-breathing volumetric liver $R_2^*$ and proton density fat
  fraction quantification in pediatric patients using stack-of-radial MRI With
  self-gating motion compensation}.
\newblock J Magn Reson Imaging 2020;53:118--129.

\bibitem[{Zhong et~al.(2020{\natexlab{b}})Zhong, Xiaodong and Armstrong, Tess
  and Nickel, Marcel D and Kannengiesser, Stephan A R and Pan, Li and Dale,
  Brian M and Deshpande, Vibhas and Kiefer, Berthold and Wu, Holden
  H}]{zhong_2020_resr2s}
Zhong X, Armstrong T, Nickel MD, Kannengiesser SAR, Pan L, Dale BM, et~al.
\newblock {Effect of respiratory motion on free-breathing 3D stack-of-radial
  liver $R_2^*$ relaxometry and improved quantification accuracy using
  self-gating}.
\newblock Magn Reson Med 2020;83:1964--1978.

\bibitem[{Fessler and Sutton(2003)Fessler, Jeffrey A and Sutton, Bradley
  P}]{fessler_2003_nufft}
Fessler JA, Sutton BP.
\newblock {Nonuniform fast Fourier transforms using min-max interpolation}.
\newblock IEEE Trans Signal Process 2003;51:560--574.

\bibitem[{Yu et~al.(2007)Yu, Huanzhou and McKenzie, Charles A and Shimakawa,
  Ann and Vu, Anthony T and Brau, Anja C S and Beatty, Philip J and Pineda,
  Angel R and Brittain, Jean H and Reeder, Scott B}]{yu_2007_t2sideal}
Yu H, McKenzie CA, Shimakawa A, Vu AT, Brau ACS, Beatty PJ, et~al.
\newblock {Multiecho reconstruction for simultaneous water-fat decomposition
  and $T_2^*$ estimation}.
\newblock J Magn Reson Imaging 2007;26:1153--1161.

\bibitem[{Yu et~al.(2008)Yu, Huanzhou and Shimakawa, Ann and McKenzie, Charles
  A and Brodsky, Ethan and Brittain, Jean H and Reeder, Scott
  B}]{yu_2008_mft2sideal}
Yu H, Shimakawa A, McKenzie CA, Brodsky E, Brittain JH, Reeder SB.
\newblock {Multiecho water-fat separation and simultaneous $R_2^*$ estimation
  with multifrequency fat spectrum modeling}.
\newblock Magn Reson Med 2008;60:1122--1134.

\bibitem[{Chebrolu et~al.(2010)Chebrolu, Venkata V and Hines, Catherine D G and
  Yu, Huanzhou and Pineda, Angel R and Shimakawa, Ann and McKenzie, Charles A
  and Samsonov, Alexey and Brittain, Jean H and Reeder, Scott
  B}]{chebrolu_2010_indiwf}
Chebrolu VV, Hines CDG, Yu H, Pineda AR, Shimakawa A, McKenzie CA, et~al.
\newblock {Independent estimation of $T_2^*$ for water and fat for improved
  accuracy of fat quantification}.
\newblock Magn Reson Med 2010;63:849--857.

\bibitem[{Reeder et~al.(2005)Reeder, Scott B and Pineda, Angel R and Wen,
  Zhifei and Shimakawa, Ann and Yu, Huanzhou and Brittain, Jean H and Gold,
  Garry E and Beaulieu, Christopher H and Pelc, Norbert J}]{reeder_2005_ideal}
Reeder SB, Pineda AR, Wen Z, Shimakawa A, Yu H, Brittain JH, et~al.
\newblock {Iterative decomposition of water and fat with echo asymmetry and
  least-squares estimation (IDEAL): Application with fast spin-echo imaging}.
\newblock Magn Reson Med 2005;54:636--644.

\bibitem[{Hernando et~al.(2010)Hernando, Diego and Kellman, Peter and Haldar,
  Justin P and Liang, Zhi-Pei}]{hernando_2010_gc}
Hernando D, Kellman P, Haldar JP, Liang ZP.
\newblock {Robust water/fat separation in the presence of large field
  inhomogeneities using a graph cut algorithm}.
\newblock Magn Reson Med 2010;63:79--90.

\bibitem[{Zhong et~al.(2014)Zhong, Xiaodong and Nickel, Marcel D and
  Kannengiesser, Stephan A R and Dale, Brian M and Kiefer, Berthold and Bashir,
  Mustafa R}]{zhong_2014_wfadafit}
Zhong X, Nickel MD, Kannengiesser SAR, Dale BM, Kiefer B, Bashir MR.
\newblock {Liver fat quantification using a multi-step adaptive fitting
  approach with multi-echo GRE imaging}.
\newblock Magn Reson Med 2014;72:1353--1365.

\bibitem[{Schneider et~al.(2020)Schneider, Manuel and Benkert, Thomas and
  Solomon, Eddy and Nickel, Dominik and Fenchel, Matthias and Kiefer, Berthold
  and Maier, Andreas and Chandarana, Hersh and Block, Kai
  Tobias}]{schneider_2020_mobawfr2s}
Schneider M, Benkert T, Solomon E, Nickel D, Fenchel M, Kiefer B, et~al.
\newblock {Free-breathing fat and $R_2^*$ quantification in the liver using a
  stack-of-stars multi-echo acquisition with respiratory-resolved model-based
  reconstruction}.
\newblock Magn Reson Med 2020;84:2592--2605.

\bibitem[{Block et~al.(2009)Block, Kai Tobias and Uecker, Martin and Frahm,
  Jens}]{block_2009_mobat2}
Block KT, Uecker M, Frahm J.
\newblock {Model-based iterative reconstruction for radial fast spin-echo MRI}.
\newblock IEEE Trans Med Imaging 2009;28:1759--1769.

\bibitem[{Fessler(2010)Fessler, Jeff A}]{fessler_2010_moba}
Fessler JA.
\newblock {Model-based image reconstruction for MRI}.
\newblock IEEE Signal Processing Magazine 2010;27:81--89.

\bibitem[{Doneva et~al.(2010)Doneva, Mariya and B\"{o}rnert, Peter and Eggers,
  Holger and Mertins, Alfred and Pauly, John and Lustig,
  Michael}]{doneva_2010_mobawf}
Doneva M, B\"{o}rnert P, Eggers H, Mertins A, Pauly J, Lustig M.
\newblock Compressed sensing for chemical shift-based water-fat separation.
\newblock Magn Reson Med 2010;64:1749--1759.

\bibitem[{Wang et~al.(2022)Wang, Nan and Cao, Tianle and Han, Fei and Xie,
  Yibin and Zhong, Xiaodong and Ma, Sen and Kwan, Alan and Fan, Zhaoyang and
  Han, Hui and Bi, Xiaoming and Noureddin, Mazen and Deshpande, Vibhas and
  Christodoulou, Anthony G. and Li, Debiao}]{wang_2022_mt-me}
Wang N, Cao T, Han F, Xie Y, Zhong X, Ma S, et~al.
\newblock Free-breathing multitasking multi-echo MRI for whole-liver
  water-specific $T_1$, proton density fat fraction, and $R_2^*$
  quantification.
\newblock Magn Reson Med 2022;87:120--137.

\bibitem[{Christodoulou et~al.(2018)Christodoulou, Anthony G and Shaw, Jaime L
  and Nguyen, Christopher and Yang, Qi and Xie, Yibin and Wang, Nan and Li,
  Debiao}]{christodoulou_2018_mt}
Christodoulou AG, Shaw JL, Nguyen C, Yang Q, Xie Y, Wang N, et~al.
\newblock {Magnetic resonance multitasking for motion-resolved quantitative
  cardiovascular imaging}.
\newblock Nat Biomed Eng 2018;2:215--226.

\bibitem[{Starekova et~al.(2022)Starekova, Jitka and Zhao, Ruiyang and Colgan,
  Timothy J and Johnson, Kevin M and Rehm, Jennifer L and Wells, Shane A and
  Reeder, Scott B and Hernando, Diego}]{starekova_2022_fb-fat}
Starekova J, Zhao R, Colgan TJ, Johnson KM, Rehm JL, Wells SA, et~al.
\newblock Improved free-breathing liver fat and iron quantification using a 2D
  chemical shift-encoded MRI with flip angle modulation and motion-corrected
  averaging.
\newblock Eur Raiol 2022;xx:1--12.

\bibitem[{Buades et~al.(2005)Buades, A. and Coll, B. and Morel,
  J.-M.}]{buades_2005_nlm}
Buades A, Coll B, Morel JM.
\newblock A non-local algorithm for image denoising.
\newblock In: Proc. IEEE Comput. Soc. Conf. Comput. Vis. Pattern Recognit.,
  vol.~2; 2005. p. 60--65.

\bibitem[{Sutton et~al.(2004)Sutton, Bradley P and Noll, Douglas C and Fessler,
  Jeffrey A}]{sutton_2004_dynamicfield}
Sutton BP, Noll DC, Fessler JA.
\newblock {Dynamic field map estimation using a spiral-in/spiral-out
  acquisition}.
\newblock Magn Reson Med 2004;51:1194--1204.

\bibitem[{Olafsson et~al.(2008)Olafsson, Valur T and Noll, Douglas C and
  Fessler, Jeffrey A}]{olafsson_2008_joint}
Olafsson VT, Noll DC, Fessler JA.
\newblock {Fast joint reconstrution of dynamic $R_2^*$ and field maps in
  functional MRI}.
\newblock IEEE Trans Med Imaging 2008;27:1177--1188.

\bibitem[{Funai et~al.(2008)Funai, Amanda K and Fessler, Jeffrey A and Yeo,
  Desmond T B and Olafsson, Valur T and Noll, Douglas
  C}]{funai_2008_secondorder}
Funai AK, Fessler JA, Yeo DTB, Olafsson VT, Noll DC.
\newblock {Regularized field map estimation in MRI}.
\newblock IEEE Trans Med Imaging 2008;27:1484--1494.

\bibitem[{Block et~al.(2014)Block, Kai Tobias and Chandarana, Hersh and Milla,
  Sarah and Bruno, Mary and Mulholland, Tom and Fatterpekar, Girish and
  Hagiwara, Mari and Grimm, Robert and Geppert, Christian and Kiefer, Berthold
  and Sodickson, Daniel K}]{block_2014_rad}
Block KT, Chandarana H, Milla S, Bruno M, Mulholland T, Fatterpekar G, et~al.
\newblock {Towards routine clinical use of radial stack-of-stars 3D
  gradient-echo sequences for reducing motion sensitivity}.
\newblock J Korean Soc Magn Reson Med 2014;18:87--106.

\bibitem[{Uecker et~al.(2008)Uecker, Martin and Hohage, Thorsten and Block, Kai
  Tobias and Frahm, Jens}]{uecker_2008_nlinv}
Uecker M, Hohage T, Block KT, Frahm J.
\newblock {Image reconstruction by regularized nonlinear inversion -- Joint
  estimation of coil sensitivities and image content}.
\newblock Magn Reson Med 2008;60:674--682.

\bibitem[{Boyd et~al.(2010)Boyd, Stephen and Parikh, Neal and Chu, Eric and
  Peleato, Borja and Eckstein, Jonathan}]{boyd_2010_admm}
Boyd S, Parikh N, Chu E, Peleato B, Eckstein J.
\newblock {Distributed optimization and statistical learning via the
  alternating direction method of multipliers}.
\newblock Foundations and Trends in Machine Learning 2010;3:1--122.

\bibitem[{Tan et~al.(2019)Tan, Zhengguo and Voit, Dirk and Kollmeier, Jost M
  and Uecker, Martin and Frahm, Jens}]{tan_2019_mobawf}
Tan Z, Voit D, Kollmeier JM, Uecker M, Frahm J.
\newblock {Dynamic water/fat separation and $B_0$ inhomogeneity mapping --
  Joint estimation using undersampled triple-echo multi-spoke radial FLASH}.
\newblock Magn Reson Med 2019;82:1000--1011.

\bibitem[{Winkelmann et~al.(2007)Winkelmann, Stefanie and Schaeffter, Tobias
  and Koehler, Thomas and Eggers, Holger and Doessel,
  Olaf}]{winkelmann_2007_golden}
Winkelmann S, Schaeffter T, Koehler T, Eggers H, Doessel O.
\newblock {An optimal radial profile based on the golden ratio for
  time-resolved MRI}.
\newblock IEEE Trans Med Imaging 2007;26:68--76.

\bibitem[{Untenberger et~al.(2016)Untenberger, Markus and Tan, Zhengguo and
  Voit, Dirk and Joseph, Arun A and Roeloffs, Volkert and Merboldt,
  Klaus-Dietmar and Sch\"{a}tz, Sebastian and Frahm,
  Jens}]{untenberger_2016_asym}
Untenberger M, Tan Z, Voit D, Joseph AA, Roeloffs V, Merboldt KD, et~al.
\newblock {Advances in real-time phase-contrast flow MRI using asymmetric
  radial gradient echoes}.
\newblock Magn Reson Med 2016;75:1901--1908.

\bibitem[{Roemer et~al.(1990)Roemer, P B and Edelstein, W A and Hayes, C E and
  Souza, S P and Mueller, O M}]{roemer_1990_pi}
Roemer PB, Edelstein WA, Hayes CE, Souza SP, Mueller OM.
\newblock {The NMR phased array}.
\newblock Magn Reson Med 1990;16:192--225.

\bibitem[{Pruessmann et~al.(1999)Pruessmann, Klaas P and Weiger, Markus and
  Scheidegger, Markus B and Boesiger, Peter}]{pruessmann_1999_sense}
Pruessmann KP, Weiger M, Scheidegger MB, Boesiger P.
\newblock {SENSE: Sensitivity encoding for fast MRI}.
\newblock Magn Reson Med 1999;42:952--962.

\bibitem[{Griswold et~al.(2002)Griswold, Mark A and Jakob, Peter M and
  Heidemann, Robin M and Nittka, Mathias and Jellus, Vladimir and Wang, Jianmin
  and Kiefer, Berthold and Haase, Axel}]{griswold_2002_grappa}
Griswold MA, Jakob PM, Heidemann RM, Nittka M, Jellus V, Wang J, et~al.
\newblock {Generalized autocalibrating partially parallel acquisitions
  (GRAPPA)}.
\newblock Magn Reson Med 2002;47:1202--1210.

\bibitem[{Reeder et~al.(2012)Reeder, Scott B and Bice, Emily K and Yu, Huanzhou
  and Hernando, Diego and Pineda, Angel}]{reeder_2012_perfr2s}
Reeder SB, Bice EK, Yu H, Hernando D, Pineda A.
\newblock {On the performance of $T_2^*$ correction methods for quantification
  of hepatic fat content}.
\newblock Magn Reson Med 2012;67:389--404.

\bibitem[{Pruessmann et~al.(2001)Pruessmann, Klaas P and Weiger, Markus and
  B\"{o}rnert, Peter and Boesiger, Peter}]{pruessmann_2001_gsense}
Pruessmann KP, Weiger M, B\"{o}rnert P, Boesiger P.
\newblock {Adcances in sensitivity encoding with arbitrary k-space
  trajectories}.
\newblock Magn Reson Med 2001;46:638--651.

\bibitem[{Feng et~al.(2014)Feng, Li and Grimm, Robert and Block, Kai Tobias and
  Chandarana, Hersh and Kim, Sungheon and Xu, Jian and Axel, Leon and
  Sodickson, Daniel K and Otazo, Rocardo}]{feng_2014_grasp}
Feng L, Grimm R, Block KT, Chandarana H, Kim S, Xu J, et~al.
\newblock {Golden-angle radial sparse parallel MRI: Combination of compressed
  sensing, parallel imaging, and golden-angle radial sampling for fast and
  flexible dynamic volumetric MRI}.
\newblock Magn Reson Med 2014;72:707--717.

\bibitem[{Hines et~al.(2009)Hines, CDG and Yu, Huanzhou and Shimakawa, Ann and
  McKenzie, Charles A and Brittain, Jean H and Reeder, Scott
  B}]{hines_2009_wfiron}
Hines C, Yu H, Shimakawa A, McKenzie CA, Brittain JH, Reeder SB.
\newblock {$T_1$ independent, $T_2^*$ corrected MRI with accurate spectral
  modeling for quantification of fat: Validation in a fat-water-SPIO phantom}.
\newblock J Magn Reson Imaging 2009;30:1215--1222.

\bibitem[{Bush et~al.(2018)Bush, Emily C and Gifford, Aliya and Coolbaugh,
  Crystal L and Towse, Theodore F and Damon, Bruce M and Welch, E
  Brian}]{bush_2018_fat}
Bush EC, Gifford A, Coolbaugh CL, Towse TF, Damon BM, Welch EB.
\newblock {Fat-water phantoms for magnetic resonance imaging validation: A
  flexible and scalable protocol}.
\newblock J VIS EXP 2018;139:1--9.

\bibitem[{Uecker et~al.(2015)Uecker, Martin and Ong, Frank and Tamir, Jonathan
  I and Bahri, Dara and Virtue, Patrick and Cheng, Joseph Y and Zhang, Tao and
  Lustig, Michael}]{uecker_2015_bart}
Uecker M, Ong F, Tamir JI, Bahri D, Virtue P, Cheng JY, et~al.
\newblock {Berkeley Advanced Reconstruction Toolbox}.
\newblock In: Proceedings of the 23th Annual Meeting of ISMRM, Toronto, CAN;
  2015. p. 2486.

\bibitem[{Huang et~al.(2008)Huang, Feng and Vijayakumar, Sathya and Li, Yu and
  Hertel, Sarah and Duensing, George R}]{huang_2008_scc}
Huang F, Vijayakumar S, Li Y, Hertel S, Duensing GR.
\newblock {A software channel compression technique for faster reconstruction
  with many channels}.
\newblock Magn Reson Imaging 2008;26:133--141.

\bibitem[{Rosenzweig et~al.(2019)Rosenzweig, Sebastian and Holme, H Christian M
  and Uecker, Martin}]{rosenzweig_2019_ring}
Rosenzweig S, Holme HCM, Uecker M.
\newblock {Simple auto-calibrated gradient delay estimation from few spokes
  using Radial Intersections (RING)}.
\newblock Magn Reson Med 2019;81:1898--1906.

\bibitem[{Rosenzweig et~al.(2020)Rosenzweig, Sebastian and Scholand, Nick and
  Holme, H Christian M and Uecker, Martin}]{rosenzweig_2020_ssa}
Rosenzweig S, Scholand N, Holme HCM, Uecker M.
\newblock {Cardiac and Respiratory Self-Gating in Radial MRI using an Adapted
  Singular Spectrum Analysis (SSA-FARY)}.
\newblock IEEE Trans Med Imaging 2020;39:3029--3041.

\bibitem[{Liu et~al.(2007)Liu, Chia-Ying and McKenzie, Charles A and Yu,
  Huanzhou and Brittain, Jean H and Reeder, Scott B}]{liu_2007_ff}
Liu CY, McKenzie CA, Yu H, Brittain JH, Reeder SB.
\newblock {Fat quantification with IDEAL gradient echo imaging: Correction of
  bias from T1 and noise}.
\newblock Magn Reson Med 2007;57:354--364.

\bibitem[{Uecker et~al.(2014)Uecker, Martin and Lai, Peng and Murphy, Mark J
  and Virtue, Patrick and Elad, Michael and Pauly, John M and Vasanawala,
  Shreyas S and Lustig, Michael}]{uecker_2014_espirit}
Uecker M, Lai P, Murphy MJ, Virtue P, Elad M, Pauly JM, et~al.
\newblock ESPIRiT -- an eigenvalue approach to autocalibrating parallel MRI:
  Where SENSE meets GRAPPA.
\newblock Magn Reson Med 2014;71:990--1001.

\bibitem[{Zhang et~al.(2021)Zhang, Guanglu and Allaire, Douglas and Cagan,
  Jonathan}]{zhang_2021_nlls_init}
Zhang G, Allaire D, Cagan J.
\newblock {Taking the guess work out of the initial guess: A solution interval
  method for least-squares parameter estimation in nonlinear models}.
\newblock ASME J Comput Inf Sci Eng 2021;21.
\newblock 021011.

\bibitem[{Hu et~al.(2020)Hu, Yuxin and Wang, Xiaole and Tian, Qiyuan and Yang,
  Grant and Daniel, Bruce and McNab, Jennifer and Hargreaves,
  Brian}]{hu_2020_spa_llr}
Hu Y, Wang X, Tian Q, Yang G, Daniel B, McNab J, et~al.
\newblock {Multi-shot diffusion-weighted MRI reconstruction with
  magnitude-based spatial-angular locally low-rank regularization (SPA-LLR)}.
\newblock Magn Reson Med 2020;83:1596--1607.

\bibitem[{Tan et~al.(2017)Tan, Zhengguo and Hohage, Thorsten and Kalentev,
  Olkesandr and Joseph, Arun A and Wang, Xiaoqing and Voit, Dirk and Merboldt,
  Klaus-Dietmar and Frahm, Jens}]{tan_2017_scalemobaflow}
Tan Z, Hohage T, Kalentev O, Joseph AA, Wang X, Voit D, et~al.
\newblock {An eigenvalue approach for the automatic scaling of unknowns in
  model-based reconstructions: Applications to real-time phase-contrast flow
  MRI}.
\newblock NMR Biomed 2017;30:e3835.

\bibitem[{Breuer et~al.(2005)Breuer, Felix A and Blaimer, Martin and Heidemann,
  Robin M and Mueller, Matthias F and Griswold, Mark A and Jakob, Peter
  M}]{breuer_2005_caipi}
Breuer FA, Blaimer M, Heidemann RM, Mueller MF, Griswold MA, Jakob PM.
\newblock {Controlled Aliasing in Parallel Imaging Results in Higher
  Acceleration (CAIPIRINHA) for Multi-Slice Imaging}.
\newblock Magn Reson Med 2005;53:684--691.

\bibitem[{Solomon et~al.(2021)Solomon, Eddy and Rigie, David S. and Vahle,
  Thomas and Paška, Jan and Bollenbeck, Jan and Sodickson, Daniel K. and
  Boada, Fernando E. and Block, Kai Tobias and Chandarana,
  Hersh}]{solomon_2021_pilot_tone}
Solomon E, Rigie DS, Vahle T, Paška J, Bollenbeck J, Sodickson DK, et~al.
\newblock Free-breathing radial imaging using a pilot-tone radiofrequency
  transmitter for detection of respiratory motion.
\newblock Magnetic Resonance in Medicine 2021;85:2672--2685.

\end{thebibliography}

\end{document}